\documentclass[useAMS,usegraphicx,usenatbib,11pt]{mn2e}
\usepackage{bm}
\usepackage[fleqn]{amsmath}
\usepackage{amssymb}
\usepackage[english]{babel}  
\usepackage{array}
\usepackage{graphicx}
\usepackage{multicol}
\usepackage{times}
\usepackage{longtable}
\usepackage{soul}

\newcommand{\rhrj}{\mathcal{R}_{\rm hJ}} 
\newcommand{\rcrh}{\mathcal{R}_{\rm ch}} 

\newcommand{\rhrjr}{\mathcal{R}_{1}}

\newcommand{\rhrjo}{\mathcal{R}_0}
 
\newcommand{\rv}{r_{\rm v}}
\newcommand{\rc}{r_{\rm c}}
\newcommand{\rj}{r_{\rm J}}
\newcommand{\gc}{\gamma_{\rm c}}
\newcommand{\dr}{{\rm d}}
\newcommand{\msun}{{\rm M}_\odot}
\newcommand{\rg}{R_{\rm G}}
\newcommand{\rp}{R_{\rm P}}
\newcommand{\ra}{R_{\rm A}}
\newcommand{\mg}{M_{\rm G}}

\newcommand{\vg}{V_{\rm G}}

\newcommand{\trh}{\tau_{\rm rh}}

\newcommand{\trhi}{\tau_{\rm rh0}}

\newcommand{\mlo}{m_{\rm low}}
\newcommand{\mup}{m_{\rm up}}
\newcommand{\mupi}{m_{\rm up}^{\infty}}

\newcommand{\kms}{\mbox{km\,s}^{-1}}
\newcommand{\rh}{r_{\rm h}} 
\newcommand{\rhdot}{\dot{r}_{\rm h}}
\newcommand{\rhi}{r_{\rm h0}} 
\newcommand{\mm}{\bar{m}}
\newcommand{\ms}{\mathcal{M}}
\newcommand{\msd}{\dot{\mathcal{M}}}
\newcommand{\msi}{\mathcal{M}_{\rm 1}}
\newcommand{\gev}{\gamma_{\rm s}}
\newcommand{\gdis}{\gamma_{\rm e}}
\newcommand{\mmse}{\bar{m}_{\rm s}}

\newcommand{\mmsed}{{\dot{\bar{m}}_{\rm s}}}

\newcommand{\tdyn}{\tau_{\rm b}}

\newcommand{\tse}{\tau_{\rm e}}

\newcommand{\find}{{f}_{\rm ind}}

\newcommand{\Msun}{\;M_{\odot}}
\newcommand{\F}{\mathcal{F}}
\newcommand{\X}{\mathcal{X}}
\newcommand{\Y}{\mathcal{Y}}
\newcommand{\U}{\mathcal{U}}
\newcommand{\Seg}{\mathcal{S}}
\newcommand{\Pa}{\mathcal{P}}

\newcommand{\xie}{\xi_{\rm e}}
\newcommand{\xii}{\xi_{\rm i}}

\newcommand{\me}{m_{\rm esc}}

\newcommand{\pr}{\psi_{\rm 1}}
\newcommand{\po}{\psi_{\rm 0}}

\newcommand{\nc}{n_{\rm c}}

\usepackage{color}

\title[A fast code for evolution of star clusters -- III]{A prescription and fast code for the long-term evolution of star clusters -- III. Unequal masses and stellar evolution}
\author[P.~E.~R.~Alexander et al.]{Poul~E.R. Alexander$^{1}$\thanks{e-mail: pera@ast.cam.ac.uk}, Mark Gieles$^{2}$, Henny~J.G.L.M. Lamers$^{3}$, and Holger~Baumgardt$^4$\\ 
$^1$Institute of Astronomy, University of Cambridge, Madingley Road, Cambridge, CB3 0HA, UK\\
$^2$Department of Physics, University of Surrey, Guildford, GU2 7XH, UK\\
$^3$Astronomical Institute Anton Pannekoek, University of Amsterdam, P.O. Box 94249, NL-1090GE Amsterdam, The Netherlands\\
$^4$School of Mathematics and Physics, University of Queensland, St. Lucia, QLD 4072, Australia\\}
\date{Accepted 2014 May 5. Received 2014 May 5; in original form: 2014 March 17}

\begin{document} 
 
\maketitle
\begin{abstract}
{We present a new version of the fast star cluster evolution code \textsc{Evolve Me A Cluster of StarS (emacss)}. While previous versions of \textsc{emacss} reproduced clusters of single-mass stars, this version models clusters with an evolving stellar content. Stellar evolution dominates early evolution, and leads to: (1) reduction of the mean mass of stars due to the mass loss of high-mass stars; (2) expansion of the half-mass radius; (3) for (nearly) Roche Volume filling clusters, the induced escape of stars. Once sufficient relaxation has occurred ($\simeq 10$ relaxation times-scales), clusters reach a second, `balanced' state whereby the core releases energy as required by the cluster as a whole. In this state: (1) stars escape due to tidal effects faster than before balanced evolution; (2) the half-mass radius expands or contracts depending on the Roche volume filling factor; and (3) the mean mass of stars increases due to the preferential ejection of low-mass stars.

We compare the \textsc{emacss} results of several cluster properties against $N$-body simulations of clusters spanning a range of initial number of stars, mass, half-mass radius, and tidal environments, and show that our prescription accurately predicts cluster evolution for this database. Finally, we consider applications for \textsc{emacss}, such as studies of galactic globular cluster populations in cosmological simulations.}

\end{abstract}

\section{Introduction}

%Introduction paragraph - N-body problem is mathematical, star clusters more complicated. Here look at SE as the main perturbation.
In this paper we study the complex dynamical interplay between star cluster (SC) evolution in a tidal field and the evolution of stars. As a result of this interaction, SC evolution differs from that of an idealised (single-mass) cluster, leading to markedly different mass, radius, and stellar mass function \citep[MF; e.g.][]{CW1990, FH1995, LBG2010,LBG2013,WMVPZ2013}. Our objectives for this study are two-fold: first, we intend to isolate and account for the most significant effects resulting from stellar evolution in SCs. Second, we will include these effects into the fast code \textsc{Evolve Me A Cluster of StarS} \citep[][hereafter Paper I and Paper II, respectively]{AG2012,GALB2014}, in order to make this fast prescription more applicable to SC population studies.

%Computation? How current models combine the two, and studies based on these. 
The interaction between the multiple physical processes driving SC evolution has historically been explored by numerical simulations. \citet{CW1990} first combined a simple stellar evolution prescription with a Fokker-Planck code to model the evolution of Galactic globular clusters. This work has been followed by further studies: \citet{GO1997,A1999,VMPZ2009}, and can now also include effects such as radiative transfer and gas hydrodynamics \citep[e.g.][and references therein]{WMVPZ2013}. Such studies are instrumental when studying the internal dynamics of individual clusters \citep[see][]{HG2008,GH2011,SH2013}. With ongoing improvements in computational power \citep[e.g. by using Graphics Processing Units (GPUs),][]{NA2012} large-scale simulations have become increasingly accessible \citep[see, for example,][]{ZKBH2011}, although remain unable to simulate the vast populations of SCs present around galaxies. 

%References paragraph - SE in star clusters, dynamics in star clusters
SCs lose mass and eventually dissolve due to a number of internal and external processes, which have been described by several previous works. Dynamically, SC evolution occurs mainly due to relaxation \citep{A1938,K1958,S1987} which incrementally diffuses energy throughout the cluster \citep{L1970}. During this process, cluster mass decreases due to the escape of stars, which is accelerated by the tidal truncation \citep{H1961,GHZ2011}. {Meanwhile, mass-loss from individual stars, either through stellar winds or supernova explosions, results in a loss of the cluster's binding energy that can drive this dynamical relaxation \citep{GBHL2010}.}

%What is analytically found, what's already in EMACSS
\citet{GHZ2011} realised that it is possible to relate the evolution of cluster radius to mass and time for clusters in a tidal field through the conduction of energy (a phase the authors describe as `balanced' evolution). In Paper I, this relationship is used to develop a prescription (\textsc{emacss} version 1) through which the evolution of SC mass and radius can be accurately recovered for a grid of clusters of single-mass stars. In that paper, the fractional change of energy per half-mass relaxation time was assumed to remain constant, and is combined with various mechanisms for mass-loss: \citet{BHH2002} show that in isolation mass is slowly lost through relaxation alone, while \citet{FH2000,B2001,GB2008} describe the rate of mass-loss in a tidal field as a function of both relaxation and crossing times. In Paper~II we extended the description to include the evolution of the core \citep[and the related `gravothermal catastrophe';][]{LBW1968} and the escape of stars in the pre-collapse phase of single-mass clusters.

The main objectives of this paper are:
\begin{enumerate}
\item{
To include mass-loss on account of stellar evolution \citep[e.g.][]{HPT2000,LBG2010}. This mass loss leads to a decrease of the mean mass of stars while the number of stars almost does not change. The corresponding change in energy causes an `unbalanced' stage of evolution for clusters with sufficiently long relaxation times.
}

\item{
To calculate the rate at which energy changes during unbalanced evolution. This will depend on \emph{where} mass is lost within the cluster. Thus, we develop a simple parametrisation for mass segregation, and the location of the highest mass surviving stars in SCs \citep[see][]{LBG2013}. 
}

\item{
To assess the unbalanced evolution of SCs with MFs. The description presented in Paper~II considers clusters of single-mass stars, which do not evolve. Since we now allow a mass spectrum, core collapse is expected to never achieve such high central densities and, {due to the segregation of mass species, to occur on a faster timescale \citep{BI1985} for the most massive stars}.  We note {however that the formation of binaries associated with core collapse is delayed, since stellar evolution provides excess energy that inflates or supports the core, and consequently negates the need for binary formation until a much longer time has passed. The pure outward diffusion of core energy that leads to the core collapse described in Paper II is not present, and core-collapse cannot therefore be modelled by the same procedure.} Because of this complication, {along with extensive stochastic difficulties due to the effect of the presence, or absence, of individual black-hole binaries \citep{H2007} we do not model the core radius. Instead, we assume in this paper} that balanced evolution starts after a certain amount of relaxation has occurred.
}

\item{
To determine the rate at which stars escape during unbalanced evolution, and the dominant mechanisms driving this escape.
}

\end{enumerate}
%Method
As in Papers I and II, we calibrate our model against a series of $N$-body simulations spanning a range of initial number of stars, mass, half-mass radius and tidal environment. For the sake of simplicity, we use clusters in spherical galaxy halo with a flat rotation curve (singular isothermal sphere), allow no primordial binary content, and apply a circular approximation for eccentric orbits (see section~\ref{s:ecc}) . Hence we reduce the number of physical processes present in SC evolution \citep{AG2012}, although note that with sufficient approximations simplified models can be useful tools for studying realistic populations \citep[e.g.][]{SKYK2013,AG2013,LBK2013}.

%Structure
The structure of this paper is as follows. First, in section~\ref{s:nb}, we describe the suite of $N$-body simulations used to guide our investigations and benchmark our code. Next, in section~\ref{s:define}, we overview the basic physics upon which our prescription is built, and define the key parameters that determine SC evolution. We then discuss the mechanisms of SC energy change in section~\ref{s:sources}, before looking at the consequences of energy change in sections~\ref{s:sinks1} and~\ref{s:sinks2}. The combined operation of the \textsc{emacss} code is then described in section~\ref{s:all}. We finally benchmark the enhanced code against $N$-body data in section~\ref{s:vali}, and summarise our conclusions in section~\ref{s:conc}.

\section{Description of $\bm{N}$-body simulations}
\label{s:nb}

For this study we use two series of $N$-body simulations from previous works - the Roche volume (RV) filling SCs in tidal fields from \citet{BM2003}, and the initially RV under-filling SCs in tidal fields from \citet{LBG2010}. We supplement these by an additional series of several new $N$-body simulations of isolated clusters. In total, we use 26 simulated SCs, of which 21 are in tidal fields and 5 are isolated. 

The simulated clusters from \citet{BM2003} have initial $N=32768$ (32k), $N=65536$ (64k), or $N=131072$ (128k), and are initially described by $W_0 = 5$ \citet{K1966} models with a \citet{K2001} initial mass function (IMF). The IMF ranges from $0.1M_{\odot} \le m \le 15M_{\odot}$, leading to a mean mass $\mm = 0.547M_{\odot}$ at the start of evolution. The clusters have no primordial binary stars, but initially retain all of their neutron stars and white dwarfs (note that due to the upper limit of the IMF black holes are not fomed in these simulations) as a compromise for the low upper limit of the IMF.

The models from \citet{LBG2010} range from $N=16384$ (16k) to $N=131072$ (128k), spaced at increasing factors of 2. The simulations are initially described by \citet{K1966} models with $W_0 = 5$, but have a \citet{K2001} IMF with an increased range $0.1M_{\odot} \le m \le 100\,M_{\odot}$, leading to a mean mass $\mm = 0.64M_{\odot}$ at the start of evolution. For these simulations there are no primordial binaries, 10 per cent of supernovae remnants (black holes and neutron stars) are retained. In these simulations, white dwarfs do not recieve a kick velocity upon formation, and are therefore retained (unless later dynamically ejected).

For each of the above simulations we assume a spherical galaxy with a flat rotation curve with $\vg = 220\,\kms$. We use RV filling clusters on both circular and eccentric orbits with $e = 0.5$. Those on circular orbits are situated at three galactocentric radii: near the galactic centre {($\rg = 2.8$ kpc)}, in the solar neighbourhood ($\rg = 8.5$ kpc) or on the outskirts of the disk ($\rg = 15$ kpc), while the clusters on eccentric orbits have apocentres {in the solar neighbourhood ($\ra = 8.5$ kpc)} and pericentres { near the galactic centre ($\rp = 2.84$ kpc), with a semi-major axis of $5.67$ kpc}. 

Our under-filling clusters are located in the solar neighbourhood such that $\rg = 8.5$ kpc. The initial $\rh$ used was between $1$ and $4$ pc for under-filling clusters, while for RV filling clusters the initial $\rh$ are set such that the tidal radius of the \citet{K1966} model is equal to the Jacobi radius, resulting in a ratio of half-mass to Jacobi radius of $\rhrj \equiv\rh/\rj= 0.19$. The RV filling clusters were evolved using the fourth order Hermite integrator \textsc{nbody4} \citep{MA1992,A1999}, accelerated by the GRAPE6 boards of Tokyo University. Meanwhile, the RV under-filling clusters were evolved with the GPU version of \textsc{nbody6}, which features an updated neighbour scheme \citep{NA2012}. Both $N$-body codes include realistic recipes for the effect of stellar and binary star evolution \citep{HPT2000,HTP2002}.

Our additional series of isolated cluster $N$-body simulations were again performed using the GPU accelerated version of \textsc{nbody6} \citep{NA2012}. We use clusters of $N=16384$ (16k) stars, with the same range of \citet{K2001} MFs as for our RV under-filling tidally limited clusters. The simulations have initial radii of $\rh = 0.773,1.324,3.586,6.150$ and 16.65 pc, chosen such that we have initial half mass relaxation times of $\trhi = 0.03,0.1,0.3,1$ and $3$ Gyr respectively. The simulations of isolated clusters were allowed to proceed for 13 Gyr, once again retain 10 per cent of supernova remnants (black holes and neutron stars). All white dwarfs are retained (no kicks) unless later dynamically ejected.

\section{Framework and Definitions}
\label{s:define}
The evolution of a SC is primarily driven by the diffusion of energy, since the cluster spends its entire life seeking to establish equipartition \citep{vH1957}. Consequently, there is a radial flux of energy in clusters \citep{H1961,H1965}, upon which we build our model. We assume that clusters remain in virial equilibrium throughout their entire lifetime, and therefore start from the usual expression for the total energy,
\begin{align}
E &= -\kappa\frac{GM^2}{\rh}, \notag \\
&= -\kappa\frac{G(N\mm)^2}{\rh}, \label{eq:E}
\end{align}
where $G$ is the gravitational constant, $M$ is cluster mass, $\rh$ is half-mass radius, $N$ is the total number of stars and $\mm = M/N$ denotes the mean mass of stars. The form factor $\kappa$ depends on the ratio between the virial radius $\rv$ and $\rh$ and hence depends on the density profile. This definition of $E$ is sometimes referred to as the `external' energy of a cluster, since the energy stored `internally' in binaries and multiples is not considered in this definition \citep{GH1997}. Differentiating equation~(\ref{eq:E}) with respect to time and dividing by $|E|$ (note that $E$ is always negative) we obtain
\begin{align}
\frac{\dot{E}}{|E|} = \frac{\rhdot}{\rh}-\frac{\dot{\kappa}}{\kappa}-2\frac{\dot{N}}{N}-2\frac{\dot{\mm}}{\mm}. \label{eq:diff}
\end{align}

\subsection{Timescales}
\label{s:time}
Before we can proceed with the model description, we need to consider the timescales for dynamical evolution. In the case of single-mass clusters (Papers~I and II) the timescale of evolution is the often cited half-mass relaxation timescale ($\trh$) from \citet{SH1971}

\begin{align}
\trh = \frac{0.138}{\ln \Lambda}\left(\frac{N\rh^3}{G\mm}\right)^{1/2}.
\label{eq:trh-singlemass}
\end{align}
Here  $\Lambda$ is the argument of the Coulomb logarithm and is proportional to $N$.
In the presence of a mass spectrum the energy diffusion is more efficient. \citet{SH1971} discuss the effect of a mass spectrum and introduce a quantity $\psi$ which depends on the shape of the stellar mass spectrum as $\psi\propto\overline{m^\beta}/\overline{m}^\beta$, with $\beta = 2.5$ for equipartition between all mass species. Their full expression of the half-mass relaxation time depends on $\psi$ as $\trh\propto\psi^{-1}$, which shows that the relaxation time of clusters with a mass spectrum is shorter than the classical timescale for single-mass clusters. Because stellar evolution reduces the width of the mass spectrum in time, the effect of a stellar MF is more important for young clusters.
\citet{GBHL2010} show that the time dependent `speed-up' in relaxation can be well approximated by an additional power-law term of time (in physical units), incorporated into $\trh$. %Because the expression for $\trh$ for single-mass clusters (equation~\ref{eq:trh-singlemass}) is so widespread
We therefore define  a `modified'  half-mass relaxation timescale as
\begin{align}
\trh' = \frac{\trh}{\psi(t)}, \label{eq:trhp}
\end{align}
where
\begin{align}
\psi(t) = \begin{cases}
            \pr, &t \leq \tse, \\
	    \displaystyle(\pr-\po)\left(\frac{t}{\tse(\mup)}\right)^{y}+\po, &t > \tse,
           \end{cases}
            \label{eq:psi}
\end{align}
which is an approximate form of $\psi$ which we adopt in order to avoid the integration of the MF required by the formula of \citet{SH1971}. In equation~(\ref{eq:psi}), $y$ is a small negative index, and $\pr$ is some factor by which the relaxation time is reduced at the start of the evolution. The final term, $\po > 1$, represents an asymptotic limit toward which $\psi$ evolves over time. If the MF were to reach zero width in a finite time (i.e. low and high mass stars are lost such that the MF becomes a delta function), we would expect $\po = 1$, which is therefore a lower limit. The normalisation time  $\tse(\mup)$ is the main sequence lifetime of the most massive star. We use an approximate analytic expression for $\tse(m)$ of the form

\begin{equation}
\tse(m) = \tse(\mup)\left[1+\frac{\ln(m/\mup)}{\ln(\mup/\mupi)} \right]^{a},
\label{eq:tse}
\end{equation}
where $a$ is a power that we set by comparison against models, and $\mupi$ is the value of $\mup(t)$ when $t$ is large. Because some remnants (in particular neutron stars and white dwarfs) are retained by a SC and are assumed to have infinite lifetime, we use $\mupi = 1.2$ $M_{\odot}$. {This represents our knowledge that \emph{some} neutron stars, black holes or white dwarfs are retained, but that this process is sufficiently random that we do not know their mass for any given cluster.} The lowest mass \emph{main sequence} star will have a lower mass than $\mupi$, which we assume to be constant $\mlo = 0.1\,M_{\odot}$ for this study. For $\mup = 100\,\msun$ we find from \citet{HPT2000} that a value of $\tse(\mup)= 3.3\,$Myr, and that the analytic prediction is well matched by equation~(\ref{eq:tse}) where $a = -2.7$. The accuracy of this approximation for other masses is shown in Fig.~\ref{f:H-fits}.

Finally, in $\trh'$ we use $\Lambda=0.02N$ for the argument of the Coulomb logarithm, as was found from $N$-body models of clusters with a mass spectrum \citep{GH1996}. 

\subsection{Dimensionless parameters}
Now that the timescales are defined we can proceed to define dimensionless parameters for the change in the quantities of equation~(\ref{eq:diff})  as a function of $\trh'$:
\begin{align}
\frac{\dot{E}}{|E|} &= \frac{\epsilon}{\trh'}, \label{eq:EdotE}\\
\frac{\rhdot}{\rh} &= \frac{\mu}{\trh'}, \\
\frac{\dot{\kappa}}{\kappa} &= \frac{\lambda}{\trh'}, \label{eq:kaps}\\
\frac{\dot{N}}{N} &= -\frac{\xi}{\trh'}, \label{eq:NdotN}\\
\frac{\dot{\mm}}{\mm} &= \frac{\gamma}{\trh'}, \label{eq:mmdotmm}
\end{align}
where in the balanced phase of evolution the conduction of energy is constant per $\trh'$ and equation~(\ref{eq:EdotE}) becomes
\begin{align}
\frac{\dot{E}}{|E|} &= \frac{\zeta}{\trh'} \label{eq:EdotEbal}
\end{align}
with $\zeta = 0.1$ (see Paper I, Paper II). Using these definitions, equation~(\ref{eq:diff}) can be written in dimensionless form,
\begin{align}
\epsilon = \mu-\lambda+2\xi-2\gamma. \label{eq:greek}
\end{align}
The definitions of the dimensionless parameters/functions ($\epsilon$, $\mu$, $\lambda$, $\xi$, $\gamma$) are explored in sections~\ref{s:sources}, \ref{s:sinks1} and~\ref{s:sinks2} {as follows.}

In section~\ref{s:sources} we first explore the changes in cluster energy (i.e. $\epsilon$). Following this we examine consequences of these changes (i.e. the evolution that we seek to model for clusters, expressed in terms of $\xi$, $\lambda$, $\mu$, $\gamma$), in sections~\ref{s:sinks1} and~\ref{s:sinks2}. We finally combine these various terms into a complete prescription for SC evolution in section~\ref{s:all}.

\subsection{Criterion for the start of balanced evolution}
\label{s:cc}

The time at which balanced evolution begins is critical for our prescription. In Paper II, the collapse depends on the core radius $\rc$, and occurs when $\rcrh = \rcrh^{\rm min}$, where $\rcrh\equiv\rc/\rh$  and $\rcrh^{\rm min}$ is the minimum value of $\rcrh$ found in the unbalanced phase (e.g. at the moment of core collapse). Here, however, we cannot use this definition since the core will not simply contract (as in the single-mass case) but can instead remain larger due to the energy released by stars evolving in the core \citep[section~\ref{s:se},][]{GH1996}.

We assume instead that the time that balanced evolution starts, $\tdyn$, is when a number of $\trh'$ have elapsed, i.e. 
\begin{equation}
\nc = \int_0^{\tdyn}\frac{{\rm d}t}{\trh'},
\label{eq:n}
\end{equation}
where $\nc$ is of order unity and will be determined later.

\subsection{Prescription for eccentric orbits}
\label{s:ecc}

A cluster on an eccentric orbit will experience a tidal field that varies with time. For sufficiently eccentric orbits, this variation of tidal field can be rapid at pericentre, and can occur on a timescale similar to (or shorter than) $\trh$. The results of rapidly varying tidal fields are two-fold: firstly, the Jacobi surface of a cluster will take a somewhat different shape and different properties, and may be distinct from a static approximation \citep{RGB2011}. Secondly, a close pericentre passage leads to rapid fluctuations in the tidal field, which in turn introduce adiabatic shocking terms, injecting additional energy into a SC \citep{W1994a,W1994b}. The consequences of such energetic injections are not fully constrained, and are not considered by \textsc{emacss}.

We therefore choose to model eccentric orbits in an approximate manner. \citet{BM2003} empirically show that an approximate lifetime for an SC on an eccentric orbit can be obtain by  considering the cluster to exist on a circular orbit at an $\rg$ defined as $\rg = \ra(1-e)$, where $\ra$ is apocentre and $e$ is eccentricity. We therefore assume that this approximation is also applicable to the mass-loss rate, and adopt the assumption $\rg = \ra(1-e)$ for clusters on eccentric orbits. The evolution of these clusters is then treated as occurring on a circular orbit at $\rg$. This assumption is tested in section~\ref{s:tidal} and appendix~\ref{ap:extras}.

\section{Changes in energy}
\label{s:sources}

 In the previous papers of this series we have examined the changes in energy occurring in the various stages of the life-cycle of single-mass clusters: the `unbalanced' (pre-core collapse) stage in Paper II, and the `balanced' (post core collapse) evolution in Paper I. In this section, we briefly overview these, before looking at the additional mechanisms through which an SCs energy changes due to the effects present in SC with an evolving MF.

\subsection{Pre-collapse core contraction}
\label{s:coll}
{In the absence of an energy source, the core of a cluster contracts so as to generate energy for the relaxation process \citep{LBE1980}. The total external energy of an isolated SC will not change during this stage of evolution, because energy is merely redistributed within the cluster. For clusters in a tidal field however, the energy increases due to the tidal stripping of stars with a small negative energy. The behaviour of single-mass clusters undergoing core contraction and gravothermal collapse is discussed in detail in Paper~II.}

% For clusters with a spectrum of stellar masses, \citet{PZM2002} show that a shallower core collapse {will occur} on the timescale of the dynamical friction of the most massive star. This short (dynamical) core collapse time naturally occurs in our model {due to} the inclusion of $\psi(t)$ into our definition of $\trh'$. {We find however that the dynamical collapse may be slowed or halted early if an additional source of energy becomes available to the cluster; for example, if $\trh$ is longer than the lifetime of massive stars (about $ 3$\,Myr, for stars of $100\,M_{\odot}$). In these situations, we expect a core collapse that takes a very different form to that discussed in Paper II. For most clusters therefore, we simply} allow a period of core contraction whereby $\epsilon=0$ until the time at which the MF begins to  change due to stellar evolution, $\tse$.

 For more compact clusters, it is possible that core collapse may occur prior to $\tse$\footnote{Though only for initially very compact clusters. For a cluster of $N=10^5$ and $\mm = 0.638\Msun$ an initial half-mass radius of 0.2 pc is required for a core collapse time $\simeq 0.1\trh < 2$ Myr.}. In these cases, clusters evolve with $\epsilon = 0 $ until core collapse, and thereupon enter balanced evolution. At $\tse$ stellar evolution will begin to generate energy, but the cluster will remain balanced with excess energy `stored' by an expanding core, {responding} as required to regulate the flow of energy. 

\subsection{Evolution of the stellar mass function}
\label{s:semf}
\subsubsection{Evolution of stars}
\label{s:se}
Stars lose mass throughout their evolution, as a result of both stellar winds and (for high-mass stars) the mass-loss during a supernova. Both these processes can be fast compared to $\trh$, and lead to an increase of cluster energy (because $E \propto -M^2/\rh$, and $\rh$ gets larger {while} $M$ becomes smaller; equation~\ref{eq:E}) on a timescale faster than relaxation.

The first stars to undergo significant mass-loss are those with the highest initial mass. To a good approximation, stellar evolution `instantaneously' removes mass from high-mass main sequence stars. These are replaced then with stellar-mass black holes, neutron stars, or leave no stellar remnant in the SC. Here, we assume that high-mass main sequence stars are replaced by remnants of mass $pm$ such that $N$ does not change, and that the gas and ejecta from stellar winds and supernovae escape. The value of $0<p<1$ will vary due to the varying ratio of remnant mass to main sequence progenitor mass for (i.e. on account of the range of stellar masses in the cluster), although we expect that $p \ll 1$ for the majority of a SC's evolution. As a consequence of a non-zero $p$, the mass-loss due to the evolution of each star is $(1-p)m$. The loss of high mass stars will cause $\mmse$ (the mean mass of stars if stellar evolution is the only mechanism through which the MF changes; see section~\ref{s:dmm}) to decrease in time, which we approximate by a power-law.
\begin{align}
\mmse = \mm_0\left(\frac{t}{\tse}\right)^{-\nu}, \label{eq:label}
\end{align}
where $\mm_0$ and $\tse$ are constants. For a \citet{K2001} IMF between $0.1\,\msun$ and $100\,\msun$ we have $\mm_0\simeq0.64\,\msun$, $\tse\simeq3.3\,$Myr and $\nu\simeq0.07$ \citep{GBHL2010}, which includes both the rate (with respect to time) at which stars are evolving off the main sequence and the mass loss due to each individual star's destruction. Differentiating equation~(\ref{eq:label}) with respect to time and dividing by $\mm$, we obtain,
\begin{align}
\frac{\mmsed}{\mm} = \frac{\gev}{\trh'.} \label{eq:2}
\end{align}
in which $\gev$ is the dimensionless change in $\mmse$ on a $\trh'$ time scale,
\begin{align}
\gev = -\frac{\nu\trh'}{t}\left(\frac{\mmse}{\mm}\right), \label{eq:gev}
\end{align}
In equation~(\ref{eq:gev}), the ratio $\mmse/\mm$ is not equal to unity due to the ejection of objects (see section~\ref{s:dmm}, below). Since the rate of stellar evolution is effectively internal to the SC's stars, this ratio is required to make $\gev$ dependent on other effects changing the mean mass. For example, if low-mass stars are preferentially lost, $\mm > \mmse$, and $\gev$ is smaller than for a cluster that has not lost low-mass stars.

From equation~(\ref{eq:E}) we see that $E$ increases if $\mm$ decreases, and we assume for the moment that $\dot{\kappa} \simeq 0$. The fractional energy change as the result of  mass loss can be written as
\begin{align}
\epsilon &= -\ms\gev. \label{eq:nrg-expand}
\end{align}
If mass-loss occurs without a preferred location, i.e. homologous with the density profile, then $\ms=3$ \citep[][]{H1980}. For more centrally concentrated mass-loss $\ms > 3$ The factor $\ms$ is sensitive to where the mass is lost and can be used as a proxy for the degree of mass segregation. For an isolated cluster ($\xie=0$), with a constant density profile ($\lambda=0$) we see from  equations~(\ref{eq:greek}) and (\ref{eq:nrg-expand}) that  $\mu = (2-\ms)\gev$. We hence see that if $\ms = 3$ (no mass segregation), the cluster expands as $\rh\propto 1/M$  \citep[e.g.][]{H1980,PZR2007,GBHL2010}, while a  mass segregated cluster with $\ms > 3$ expands faster (see detailed discussion in section~\ref{s:ms}).

As in Paper II, we refer to this kind of evolution as being `unbalanced', in that energy change is independent of energy demand in the cluster. The mechanisms for energy production differ between the two papers; in Paper II, core contraction redistributes the energy of core stars and energy only changes because of escaping stars, while here stellar evolution causes the SC energy to change. As a result, during this stage $\epsilon$ is different to (usually, but not necessarily, higher than) that required for balanced evolution \citep[see discussions in][]{H1965,GHZ2011}.

\subsubsection{Energy change due to stellar evolution}
\label{s:ms}

In section~\ref{s:se} we introduce the parameter $\ms$ which relates energy change to mass-loss, and argued that this could be used as a proxy for the degree of mass segregation. At early time, we expect this to increase as mass segregation will cause high mass stars to be found in (and evolve in) incrementally deeper potentials. As the MF evolves however:
\begin{enumerate}
\item{The mass of the most massive main sequence stars decreases and becomes comparable to the typical masses of remnants.}
\item{There are increasing numbers of stars with masses similar to the highest mass stars that remain in the cluster. {It therefore becomes statistically less likely that the high mass stars will be found in the strongest potential at the core, but will instead be spread throughout a larger volume with a lower \emph{average} potential.}}
\end{enumerate}
Consequently, there is an upper limit in $\ms$, $\msi$, toward which $\ms$ will evolve. We therefore let $\ms$ increase from $\ms = 3$ to $\ms \lesssim 10$ as energy is redistributed throughout the cluster.

Since mass segregation is a consequence of the redistribution of energy between stars in a cluster, we expect the segregation of stellar species to occur on a timescale  $\simeq \trh'$ (equation~\ref{eq:trhp}; see \citealt{S1969}, but also \citealt{PZM2002,FPZ2013}). Consequently, we assume the time taken for mass segregation to occur will be $\simeq {\rm a \; few}\times \trh'$, and hence simply evolve $\ms$ as
\begin{align}
\frac{\dot{\ms}}{\ms} = \frac{\ms_1-\ms}{\trh'}.\label{eq:MsdotMs}
\end{align}

\begin{figure*}
\centering
\includegraphics[width=170mm]{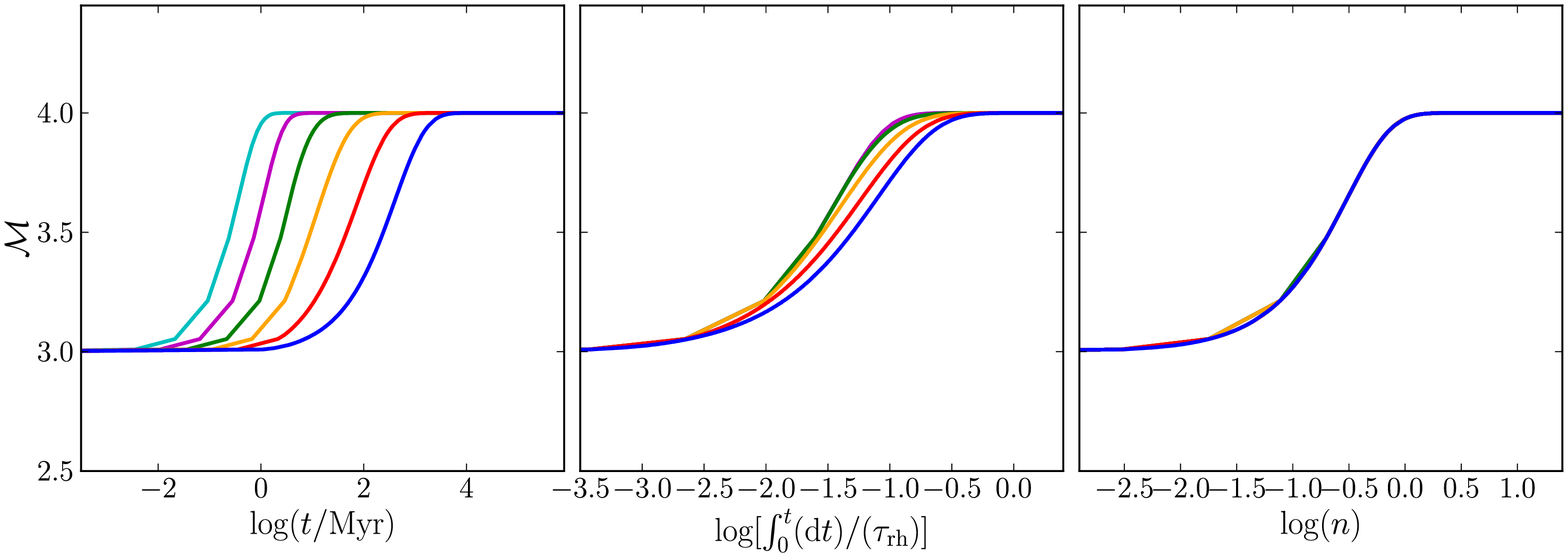}
\caption{Left panel: evolution of {the mass segregation parameter} $\ms$ as a function of $t$ for (isolated) star clusters with different $\trhi$ (left-most track (cyan) = shortest $\trhi$, right-most track (blue) = longest $\trhi$). Centre panel: evolution of $\ms$ as a function of number of elapsed relaxation $\trh$ for star clusters with the same range of $\trhi$. Right panel: evolution of $\ms$ as a function of number of elapsed modified relaxation $\trh'$ for star clusters with the same range of $\trhi$. The colours and order of lines is the same in each panel.}
\label{f:massseg}
\end{figure*}

The resulting evolution of $\ms$ is illustrated in Fig.~\ref{f:massseg}. In equation~(\ref{eq:MsdotMs}), the value of parameter $\msi$ is determined by comparison against $N$-body simulations.

\subsubsection{Ejection of low-mass stars}
\label{s:dmm}

The most significant consequence of stellar evolution is the removal of high mass stars, which causes an overall decrease of $\mm$. By contrast, stars escape across the entire range of the MF, but those that escape are preferentially of low-mass (from $t=0$, although the effect becomes more significant after mass segregation). There are two reasons for this preference: first, mass segregation forces low-mass stars into the weaker potential of the cluster halo where they are most susceptible to tidal stripping \citep{H1969,K2009,TVP2010}. Second, the typical outcome of 3 body encounters is the ejection of the least massive star, although this will only lead to escape in a minority of cases. For clusters with significant escape rates (high $\xi$), the preferential ejection of low-mass stars results in the low-mass end of the MF becoming depleted and $\mm$ increasing.

\begin{figure}
\centering
\includegraphics[height=80mm]{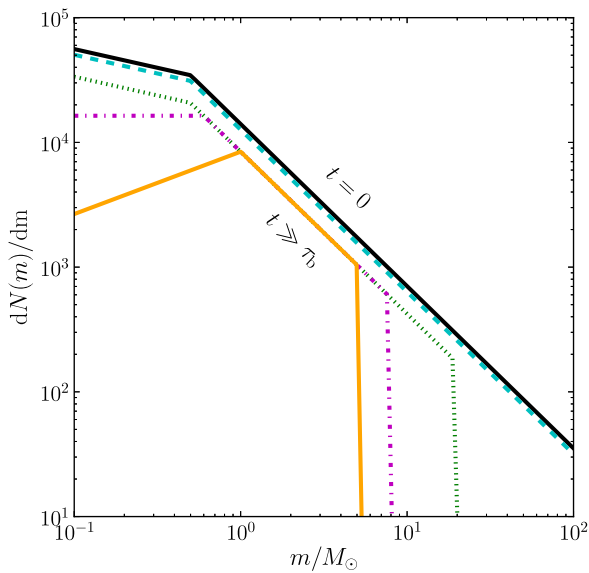}
\caption{Schematic of the changes in the MF in a $N=10^5$ cluster due to stellar evolution and the preferential ejection of low-mass stars, based upon figs. 1 and 2 of \citet{LBG2013}. The lines represent the shape of the MF at different times, from cluster formation ($t=0$) to late evolution ($t\gg\tdyn$). The initial MF (black, solid) is given by a \citet{K2001} MF between 0.1 and 100 $M_{\odot}$. Early evolution (for $t < \tse$) removes stars uniformly across the MF, without changing its shape. At $\tse$ (cyan, dashed), the highest mass stars begin to explode as supernovae, removing them from the cluster. Meanwhile stars are ejected from the entire MF until $\tdyn$ (green, dotted), whereafter we assume stars are preferentially ejected from the low-mass end of the MF. The next line (magenta, dot-dash) shows the progression of both stellar evolution and the preferential ejection of low-mass stars simultaneously, until late times when $N \ll N_0$ (orange, solid). For these last lines, the slope of the MF at the low-mass end inverts due to the absence of (remaining) low-mass stars. It is apparent that the two effects affect opposite ends of the MF.}
\label{f:masschangedemo}
\end{figure}

\begin{figure}
\centering
\includegraphics[width=80mm]{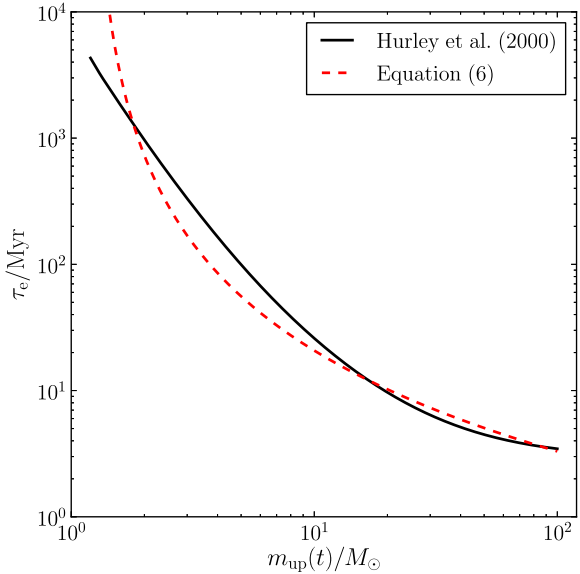}
\caption{Comparison of our analytic fit for $\tse$ (as a function of $m_{\rm up}$; see equation~\ref{eq:tse}) to the main sequence lifetimes of stars predicted by the models of \citet{HPT2000}. Note that by using our definition of $\tse$, we are able to invert the function and hence recover $m_{\rm up}(t)$ as a function of $t$. Since the assumed mass of neutron stars and white dwarfs is $m \simeq 1.2 M_{\odot}$ and we make the assumption that a fraction of these are retained by the cluster, $\tse$ is infinite for $m \leq 1.2 M_{\odot}$, (i.e. we do not expect to find $\mup(t) \geq 1.2 M_{\odot}$ at any time). For this figure, we predict $\tse$ for a range of stellar masses $1.2\,M_{\odot} \leq m \leq 100\,M_{\odot}$, with an expected main sequence lifetime of a $100\,M_{\odot}$ star $\tse = 3.3$ Myr, and use a value $a = 2.7$. We note however that \textsc{emacss} assumes a \citet{K2001} IMF with $\mlo = 0.1\,M_{\odot}$, and a value of $\mup(t)$ that defines $\tse$ as shown by this plot.}
\label{f:H-fits}
\end{figure}

Because stellar evolution and stellar ejection affect opposite ends of the MF (see Fig.~\ref{f:masschangedemo}), the effects are approximately independent and can be quantified separately \citep[see][]{LBG2013}. The overall change in $\mm$ is therefore described by
\begin{align}
\gamma = \gev+\gdis \label{eq:gammas}
\end{align}
where $\gdis$ is the change in mean mass due to the ejection of (preferentially low-mass) stars and $\gev$ is the change in mean mass due to stellar evolution (equation~\ref{eq:gev}).We define $\gdis$ as 
\begin{align}
\gdis = \left(1-\frac{\me}{\mm}\right)\Seg\U\xie, \label{eq:4}
\end{align} 
where $\me$ is the average mass of the escaping stars. Following mass segregation, this will be greater that the minimum mass of the initial MF $m_{\rm low}$, and \emph{on average} less than the mean mass $\mm$. We consider the average mass of the escaping stars to be
\begin{align}
\me = \X\left(\mm-m_{\rm low}\right)+m_{\rm low}, \label{eq:me}
\end{align} 
in which $m_{\rm low} =  0.1\,M_{\odot}$ for this study and $0\leq \X \leq 1$ is a constant. In equation~(\ref{eq:4}), the second term $\Seg$ evolves from 0 to 1 as mass segregation proceeds and is defined as 
\begin{align}
\Seg = \left(\frac{\ms-3}{\ms_1-3}\right)^q,
\end{align}
where $q$ is a power defined by comparison against $N$-body simulations, and $\ms_1$ represents the maximum value of $\ms$ found for a fully mass-segregated cluster. The final term in equation~(\ref{eq:4}),
\begin{align}
\U = \frac{m_{\rm up}(t)-\mm}{m_{\rm up}(t)} \label{eq:U}
\end{align} 
is a factor that prevents $\mm$ from growing larger than the maximum mass $m_{\rm up}(t)$ (e.g. $\U \rightarrow 0$ when $\mm \rightarrow m_{\rm up}(t)$). In order to solve equation~(\ref{eq:U}), we need to know the maximum mass of the MF at any given time. This is calculated {for} $t > \tse$ by an analytic fit to the main sequence lifetime formulae of \citet{HPT2000}, for which we find a good approximation is provided by the inverse of equation~(\ref{eq:tse}). In Fig.~\ref{f:H-fits} we demonstrate the evolution of the $m_{\rm up}(t)$ according to equation~(\ref{eq:tse}), along with the models of \citet{HPT2000}. 

Both this section and section~\ref{s:se} consider very simple descriptions for the evolution of the MF, in which the only measurable quantity is $\mm$. Although more complete descriptions for the MF exist \citep[e.g. the `differential mass function' of][]{LBG2013}, in order to retain the clarity of this work we refrain from using more complex descriptions.

\subsection{Energy of escapers during unbalanced evolution}
\label{s:sHe}

Stars escape from SCs on account of either internal processes \citep[i.e. stars are lost from isolated clusters,][]{BHH2002}, or, for clusters tidally limited by the presence of a galaxy, interaction with the tidal field \citep{FH2000}. The escape of stars causes SC mass to decrease over time, which will consequently increase the gravitational binding energy (i.e., escapers constitute a positive contribution $E$). As stars escape via the Lagrange points with $v \simeq 0$, the energy change due to each escaper will depend only on the potential experienced by the star when it escapes, i.e. $\dot{E} = \me \phi_{\rm e} \dot{N}$ and $\phi_{\rm e} = -GM/\rj$ is the potential at the Jacobi surface $\rj$.

Rearranging and dividing by $|E|$ we can obtain a contribution to $\epsilon$ due to escaping stars
\begin{align}
\epsilon = \frac{1}{\kappa}\left(\frac{\me}{\mm}\right)\left(\frac{\rh}{\rj}\right)\xi, \label{eq:eps_scape}
\end{align}
where $\xi$ is escape rate of stars; see section~\ref{s:dNub}, and Papers I and II. This contribution is only present for unbalanced clusters, as by definition balanced evolution includes this energy change. Because $\xi > 0$ (at all times), $\epsilon > 0$, and stars escaping during unbalanced evolution cause a net increase in cluster energy.

\subsection{Post-collapse (balanced) evolution}
\label{s:bal}

%Balanced paragraph - how this phase works
All clusters will eventually reach a `balanced' state \citep{GHZ2011} at the time that a regulatory mechanism is formed in the core \citep[roughly at core collapse,][]{HS2012}. In this state, energy changes within the core \citep[binary action, interactions with stellar- or intermediate- mass black holes and ongoing stellar evolution,][]{SH1971,H1975,GBHL2010,BH2013} are regulated by the behaviour of the cluster as a whole and not by local processes \citep{H1961}, leading to a radial flux of energy in the cluster that, for single-mass clusters, is constant per unit of relaxation time ($\dot{E}\trh/E=\,$constant).

For multi-mass models we find that this energy flux depends on time, such that $\dot{E}\trh/E= f(t)$ (a monotonically decreasing function of time). This decrease emanates from the decreasing ratio between $\mup(t)$ and $\mlo$, which decreases the rate at which energy can be transported through a cluster by relaxation. We have previously accounted for this decreasing efficiency of relaxation by including a $\psi$ term (equation~\ref{eq:psi}) in the definition of $\trh'$. Hence, for clusters with an evolving MF, we find that $\dot{E}\trh'/E=\,$constant, and that the constant is once again given by $\zeta$ (see equation~\ref{eq:EdotEbal}).

\section{Changes in cluster properties during unbalanced evolution}
\label{s:sinks1}

In section~\ref{s:sources}, above,  we examined the mechanisms through which the external energy of a SC changes. This change in energy is related by equation~(\ref{eq:diff}) to a number of cluster properties;  the loss of stars ($\dot{N}$), a change in cluster half mass radius ($\rhdot$), and a change in the energy form factor due to a changing density profile ($\dot{\kappa}$). We examine the evolution of each property in turn below (for unbalanced clusters).

\subsection{Escape of stars}
\label{s:dNub}

The easiest indication of SC evolution to identify is the escape of stars, which will eventually lead to the total dissolution of the cluster. \citet{LBG2010} identify two mechanisms through which stars escape - escape as a result of two body relaxation (parametrised by $\xie$), which is discussed in Papers I and II, and escape induced by stellar evolution due to the resulting expansion (induced escape; parametrised by $\xii$), discussed in \citet{LBG2010}. The overall escape rate is given by the sum of these, 
\begin{align}
\xi = \xie+\xii.
\end{align}
Both of these mechanisms occur during the unbalanced phase of cluster evolution, and are discussed below.

\subsubsection{Relaxation driven escape}

In Paper I we developed a description for the dimensionless escape rate $\xie$ using the works of \citet{GB2008} and \citet{FH2000}. From these works, we showed that $\xie \propto \rhrj^{3/2}(N/\ln \gc N)^{1/4}$ where $\rhrj \equiv \rh/\rj$ and can be understood as a RV filling factor. This term implies that $\xie$ will be higher if $\rh$ is larger compared to $\rj$. The second term $[(N/\ln\gc N)^{1/4}]$ reduces the rate of escape from low-$N$ systems, and accounts for the `escape time effect' in which escape is delayed due to the anisotropic geometry of the Jacobi surface \citep{B2001}.

In Paper II we introduce two additional terms,  $f$ and $\F$ to improve the description of $\xie$ for the unbalanced regime. The first of these terms, $f \simeq 0.3$, accounts for the lower escape rate observed in $N$-body simulations for unbalanced SCs \citep[][Paper II]{LBG2010}. The second term, $\F$ accounts for the progress of core collapse, and is defined in Paper II as $\F = \rcrh/\rcrh^{\rm min}$.

Combining these terms, we define $\xie$ in a similar manner to in Paper II,
\begin{align}
\xie = \F\xi_0(1-\Pa)+\left[f+(1-f)\F\right]\frac35\zeta\Pa, \label{eq:xi}
\end{align}
where $\xi_0$ is the intrinsic escape rate from isolated clusters and $\Pa$ is a function of $N$ and $\rhrj$. The factor $\Pa$ is defined as
\begin{align}
\Pa = \left(\frac{\mathcal{R}}{\mathcal{R}_1}\right)^z\left[\frac{N\ln(\gc N_1)}{N_1\ln(\gc N)}\right]^{1-x}. \label{eq:P}
\end{align}
where $\mathcal{R}_1$ is the scale ratio $\rh/\rj = 0.145$ \citep{H1965}, and $z = 1.61$ (Paper I). The argument of the Coulomb logarithm $\gc = 0.02$ \citep[][for clusters with a stellar MF]{S1987}, $x = 0.75$ \citep{BM2003}, and $N_1 \simeq 15000$ defines the $N$ of a cluster for which $\mathcal{R}=\mathcal{R}_1$ (Paper II). The Jacobi radius ($\rj$) required to calculate $\rhrj$ is given by
\begin{align}
\rj &= \rg^{2/3}\left(\frac{GN\mm}{2\vg^2}\right)^{1/3}, 
\label{eq:rj}
\end{align}
for a singular isothermal halo. In equation~(\ref{eq:rj}), $\rg$ is the galactocentric distance, $\vg$ is the orbital velocity of the SC around the centre of mass for the galaxy. From virial theorem, equation~(\ref{eq:rj}) can be rearranged in terms of galaxy mass using the relation $\mg = \rg\vg^2/G$.

As we can no longer apply the same definition of $\F$ as in Paper II due to the unknown {evolution of} core radius, we instead use $\F$ as a measure of the progress of core collapse in number of elapsed (modified) relaxation times.  We therefore set 
\begin{align}
\F = \begin{cases}
        0, &n < \nc/2, \\
 	\frac{2n}{\nc}-1, &\nc/2\leq n \leq, \nc \\
	1 &n> \nc,
     \end{cases}
 \label{eq:F}
\end{align}
where $n = \int_0^{t}{\rm d}t/\trh'$, and the factor of 2 is chosen so as to reproduce behaviour representative of that $\F$ in Paper II. Although the first derivative produced by this formula for $\F$ is discontinuous, using equation~(\ref{eq:F}) within our formula for $\xi$ (equation~\ref{eq:xi}; recall that $\xi$ is itself a derivative) will nonetheless result in a continuous rate of mass loss.

The values of $N_1$, $\mathcal{R}_1$ and $z$ determined in Paper I are true for clusters of single-mass stars, but do not necessarily remain useful for multi-mass clusters. We therefore redefine these quantities for clusters with MFs through comparison against $N$-body simulations, and note that they remain constant throughout both unbalanced and balanced evolution.

\subsubsection{Induced escape}
\label{s:ind}

For (nearly) RV filling clusters, \citet{LBG2010} find a second mechanism through which stars escape, on top of the direct escape described above. This `induced' loss of stars is an additional {escape term} caused by stellar evolution in the cluster. {Induced escape occurs when} stars are lost because $\rh$ increases as a result of stellar evolution whilst $\rj$ simultaneously decreases on account of the decreasing total mass (equation~\ref{eq:rj}). The dimensionless rate of induced escape is expressed as
\begin{align}
\xii = f_{\rm ind}\gev, \label{eq:ind}
\end{align}
where $f_{\rm ind}$ is a factor defining how much induced escape is caused by stellar mass loss. In \citet{LBG2010}, $f_{\rm ind}$ includes both contributions from the RV filling factor and a delay period between mass being lost by stellar evolution and the corresponding (induced) escape of stars. Here however, we use a simpler form for induced escape in which we assume the delay time is negligible and therefore that the induced escape rate depends only upon RV filling.

The factor $f_{\rm ind} \simeq 0$ for under-filling clusters, but is found to be $\simeq 1$ for RV filling clusters in which escape due to two body relaxation is negligible compared to mass loss due to stellar evolution. We adopt a definition of $f_{\rm ind}$ directly based upon the RV filling factor
\begin{align}
f_{\rm ind} = \begin{cases}
              \Y\left(\rhrj-\rhrjr\right)^b, &\rhrj > \rhrjr, \\
		0, &{\rm otherwise,}
              \end{cases}
 \label{eq:find}
\end{align}
where $b$ is a power determined by comparison against $N$-body simulations, and such that the induced mass loss $\simeq 0$ for compact clusters and grows for clusters that have $\rhrj$ greater than reference $\rhrjr$. Equation~(\ref{eq:find}) is continuous in value and first derivative (if $b \ne 1$) and is therefore sufficient for this study, although we note that the higher order derivatives of $f_{\rm ind}$ are discontinuous due to the transition of regimes at $\rhrj = \rhrjr$.  The reference $\rhrj$, $\rhrjr$, is the same for this equation as for equation~(\ref{eq:P}). The factor $\Y$ in equation~(\ref{eq:find}) defines the RV filling factor at which induced mass loss becomes significant, and is once again determined by comparison against $N$-body simulations (specifically, the $N$-body simulations of RV filling clusters).

Although equation~(\ref{eq:find}) is by construction somewhat approximate, this is unlikely to effect the application of \textsc{emacss} to realisitic globular clusters \citep[e.g.][2010 version]{H1996}. This is because the majority of clusters are likely to have formed RV under-filling \citep[see][]{EJ2012,AG2013}, where $\find \simeq 0$. Conseqeuntly, this term will only be applicable to a minority of clusters.

\subsection{Change of the density profile}
\label{s:dkub}

The original \textsc{emacss} (Paper I) correctly reproduces the evolution of virial radius ($r_{\rm v}$), but ignores the variation of $\kappa$ and therefore assumes that $\rh/r_{\rm v} = 1$ throughout SC evolution. During the balanced phase, this is a reasonable assumption as the variation in $\kappa$ is mild. During unbalanced evolution however, the variation of $\kappa$ is significant since the central density changes extremely quickly as the core undergoes collapse \citep{GH1996}. This is discussed in depth in Paper II for clusters of single-mass stars, where it is shown that $\kappa$ varies on account of a varying $\rcrh$.

In Paper II,  $\kappa$ is shown to be adequately described by an error function of $\rcrh$ which varies between $\kappa_0 \simeq 0.2$ at birth \citep[the initial value of $\kappa$ for a $W_0 = 5$][model]{K1966}, and $\kappa_1 \simeq 0.24$ at late times (roughly the value of $\kappa$ for the typical density profile measured during the balanced evolution following core collapse). This evolution of $\kappa$ corresponds to the change in energy form factor for a cluster during the `gravothermal catastrophe' \citep{LBE1980}, which is generally only experienced by clusters of single-mass stars. {For clusters containing (or containing stars sufficiently massive to evolve into) black holes, the core-radius is also linked to the retention of black holes, which is a random effect and hence cannot be reliably modelled. The evolution of half-mass radius however does not depend upon the energy sources in the core, and thus can be modelled without knowledge of the core radius and black hole retention \citep{LBK2013,BH2013}.}

In the presence of a full MF of finite range, SCs mass segregate and experience an earlier, shallower collapse \citep{GH1996}. Nonetheless, in both single-mass and multi-mass cases we find similar extrema for $\kappa$ ($\kappa_0 \simeq 0.2$ and $\kappa_1 \simeq 0.24$) {since this is predominantly a result of the overall pre- or post- collapse density profile of the entire cluster, and largely independent of the exact mechanisms of interactions within the core.} We consequently adopt a very simple form for $\lambda$ (see equation~\ref{eq:kaps}) that connects $\kappa_0$ to $\kappa_1$ during the unbalanced phase,
\begin{align}
\lambda = \begin{cases}
        0, &n < \nc/2, \\
 	\left(\kappa_1-\kappa\right)\left(\frac{2n}{\nc}-1\right), &{\rm otherwise} \\
     \end{cases} \label{eq:lam}
\end{align}
where $n = \int_0^{t}{\rm d}t/\trh'$. Note that equation~(\ref{eq:lam}) follows the same dependence on $n/\nc$ as equation~(\ref{eq:F}), since both these equations describe the progress of core collapse. The extra term in this expression $\left(\kappa_1-\kappa\right)$ means that $\kappa$ will increase on a $\trh'$ timescale from $\kappa = \kappa_0$ at $t=0$ until $\kappa = \kappa_1$ at moment of core collapse (i.e., when the number of elapsed modified relaxation times equals $\nc$).

\subsection{Evolution of the half-mass radius}

The final consequence of the changing SC energy during unbalanced evolution is the expansion (or contraction) of the half-mass radius. For (nearly) isolated clusters in which $\xi \simeq 0$, this results in an expansion $\mu \sim \ms\gev$ caused by energy released by stellar evolution. However, for clusters in tidal fields where $\xi > 0$ and $\lambda > 0$, we must rearrange equation~(\ref{eq:greek}) and instead use,
\begin{align}
\mu = \epsilon-2\xi+2\gamma+\lambda, \label{eq:mu}
\end{align}
whereby $\rh$ increases or decreases so as to equate the change in energy to the dynamical properties of the cluster. For equation~(\ref{eq:mu}) during the unbalanced phase, $\epsilon$ and $\gamma$ are defined in sections~\ref{s:coll}, \ref{s:semf} and~\ref{s:sHe}, while $\xi$, and $\lambda$ are defined in sections~\ref{s:dNub} and~\ref{s:dkub}. Depending on the values of $\epsilon$ and $2\xi+2\gamma+\lambda$, $\mu$ may be either positive or negative, and the cluster may expand or contract.

\section{Changes in cluster properties during balanced evolution}
\label{s:sinks2}

Once a SC has reached a state of balanced evolution, the energy production is controlled by the whole cluster as opposed to the energy-generating core. In this state, the energy change per $\trh'$ is given by a single quantity ($\zeta$), and does not depend on any of the mechanisms outlined in sections~\ref{s:coll}, \ref{s:semf} and~\ref{s:sHe}. Like in the unbalanced phase, the change of energy leads to several dynamical effects, although unlike in the unbalanced phase the density profile does not significantly change. We are left therefore with three evolutionary effects: the escape of stars, expansion or contraction of the half-mass radius, and the changing mean mass of stars.

\subsection{Escape of stars}
After core collapse stars continue to escape (due to relaxation, $\xie$), although in balanced evolution core collapse is complete and $\F \equiv 1$. Equation~(\ref{eq:xi}) simplifies therefore to,
\begin{align}
\xie = \xi_0(1-\Pa)+\frac35\zeta\Pa, \label{eq:xi_bal}
\end{align}
which is the same escape rate as in Paper I. We also find that there is no induced mass-loss in balanced evolution (i.e. $\xii = 0$), the expansion of $\rh$ is related\footnote{The cluster either expands toward an RV filling state, or is RV filling with roughly constant density within the Jacobi surface $\rh(t) \propto \rj(t)$} to the size of $\rj$, and hence $\xi = \xie$.

\subsection{Changing mean mass of stars}
Although the change in $\mm$ due to stellar evolution no longer defines energy production in the balanced phase, stellar evolution will continue to reduce the mass of the highest mass stars, and escaping stars will typically continue to deplete the low mass end of the MF. We find therefore that once again $\gamma = \gev+\gdis$ in balanced evolution, although this is no longer related to $\epsilon$.

\subsection{Expansion and Contraction}
As in Paper I, the cluster radius (value of $\mu$) will respond so as to balance equation~(\ref{eq:greek}). For balanced evolution in which $\epsilon = \zeta$ and $\lambda=0$, the radius therefore changes as
\begin{align}
\mu = \zeta-2\xi+2\gamma. \label{eq:mu2}
\end{align}
During balanced evolution, the value of $\mu$ can be both positive or negative (depending mainly on $\rhrj$ and therefore on the relative values of $\xi$ to $\zeta$ and $\gamma$), and hence both expansion (if $\mu > 0$) and contraction (if $\mu < 0$) are possible \citep[][Paper I]{GHZ2011}.

\section{Combined operation of \textsc{emacss}}
\label{s:all}

\renewcommand{\arraystretch}{1.4}
\begin{table*}
\centering
\caption{The definitions of $\epsilon$ during the three phases of cluster evolution. The quantity $\nc$ defines a number of elapsed modified relaxation times at which balanced evolution begins (which occurs when $t = \tdyn$), and is determined by comparison with $N$-body simulations. The remaining quantities use are defined in tables~\ref{t:parasum} and~\ref{t:parascal}.}
\begin{tabular}{ m{2cm} m{4cm} m{10cm} }

  $t<\tse$& $\displaystyle{\epsilon=\frac{1}{\kappa}\frac{\me}{\mm}\frac{\rh}{\rj}\xi}$&  Unless balance begins before the supernovae of the most massive stars, the energy of a SC does not change by any internal processes in the first few ${\rm Myr}$ (depending on the upper limit of the MF). It may however change due to external processes, such as the direct removal of stars by an external tidal field. We include this contribution, although note that it is negligible for clusters that are initially sufficiently compact.\\

  $\tse<t<\tdyn$& $\displaystyle{\epsilon=\frac{1}{\kappa}\frac{\me}{\mm}\frac{\rh}{\rj}\xi+\ms\gev}$& Unbalanced evolution, described in section~\ref{s:se}. The first term accounts for the energy change due to stars escaping directly by dynamical processes. Meanwhile, the second term represents the decrease in $\mm$ and hence increase in cluster energy through stellar evolution (which are related via equation~\ref{eq:diff}). The extent of this energy change depends on whether the mass loss occurs centrally or uniformly (i.e. depends upon the degree of mass segregation, $\ms$).\\

    $t>\tdyn$& $\displaystyle{\epsilon=\zeta}$& Balanced evolution, defined in Paper I; $\dot{E}/E\trh = {\rm const}$ for clusters of single-mass stars, but is time dependent for multi-mass clusters due to the decreasing range of the MF. In this paper, $\epsilon$ is related to $\trh'$ such that $\dot{E}/E\trh' = \zeta$, and the time dependence due to an evolving MF is incorporated into our definition of $\trh'$. This state continues until very near the final dissolution of the cluster when $N \simeq 200$ and balanced evolution breaks down because $\trh$ becomes comparable to the crossing time.
\end{tabular}
\label{t:epsilon}
\end{table*}
\renewcommand{\arraystretch}{1.0}

\begin{figure}
\centering
\includegraphics[height=90mm]{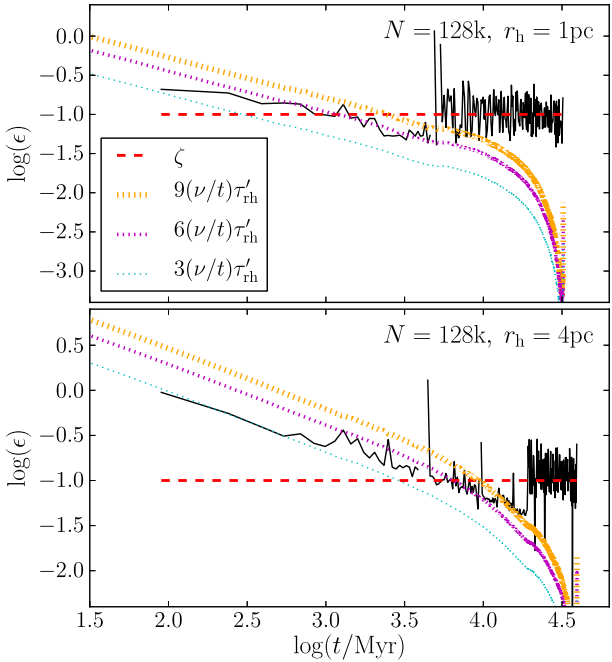}
\caption{Evolution of $\epsilon$ as a function of time for the $\rh = 1$ pc and $\rh = 4$ pc$, N=128$k SCs. In each plot,the (black) solid line is calculated using equation~(\ref{eq:EdotE}) from $N$-body data. The (red) dashed line is $\epsilon = 0.1$ (i.e. balanced evolution), while the three dotted lines are the energy changes due to stellar evolution for different (constant) mass segregation factors. Core collapse is evident as a jump in $\epsilon$. The $N$-body data crosses several of the dotted lines due to increasing mass segregation. We do not see a period without internal energy changes as the $N$-body simulations do not have sufficient resolution in the first $\simeq 100$ Myr for this to be apparent.}
\label{f:eps}
\end{figure}

Following the philosophy outlined in Papers I and II, we encapsulate the entire evolution of clusters in terms of several key quantities. Energy changes are categorised by a single parameter, $\epsilon$, which takes one of three forms at each stage of evolution: unbalanced evolution before and after $\tse$, and balanced evolution. The equations used to calculate the changing energy in each of these regimes are summarised in table~\ref{t:epsilon}, while the energy change due to the impact of these three processes is illustrated in Fig.~\ref{f:eps}.

Based on Fig.~\ref{f:eps}, we find that $\zeta = 0.1$ (as in Paper II) fits our data when $\pr = 8.0$ and $\po = 1.6$. This implies that even during the late stages of balanced evolution, the conduction of energy is more efficient for an SC with an MF than an equal-mass cluster. as previously suggested \citep[e.g.][]{LBG2013}, this result is consistent with clusters' MFs evolving toward a narrow (delta function like) shape similar to that of single-mass clusters. However, such clusters retain some range of MF for their entire lifespan. We consequently adopt these values, such that $\pr=8.0$, $\po=1.6$ and $\zeta = 0.1$ hereafter.

The transition between the second and third stages of energy production (the start of balanced evolution) occurs after a time $\tdyn$ whose physical value is not known at $t=0$. The transition between the first and second stages occurs when the highest mass stars present explode as supernova (at $\tse = 3.3\;{\rm Myr}$ for $\mup = 100 M_{\odot}$ stars), unless the SC is undergoing balanced evolution before this time.

Whereas the energy changes in a SC are summarised in table~\ref{t:epsilon}, the consequences of changing the energy of a SC are summarised in table~\ref{t:parasum}. We also define the dimensionless parameters involved in SC evolution, and the ancillary equations used by our model. Finally, table~\ref{t:parascal} summaries the key scaling factors and constants present. When known, the values of these parameters are quoted from literature.

\renewcommand{\arraystretch}{1.4}
\begin{table*}
\centering
\caption{The main equations used by \textsc{emacss}. The first section defines the differential equations that calculate the evolution of the measurable quantities (e.g. $N$, $\mm$, and $\rh$). The second section defines the dimensionless factors used in these equations, while the third section defines additional factors used by the preceding equations in each time step. Additional parameters derived from the variables modelled by \textsc{emacss} are defined in the final section.}
\begin{tabular}{r@{\;=\;}p{5cm} m{11cm} }
\hline
  \multicolumn{3}{c}{Output properties evolved by \textsc{emacss}}\\
\hline
$\dot{E}$&$\frac{\epsilon |E|}{\trh'}$  & Rate of change of the external energy of stars. Virial equilibrium is assumed throughout, such that $E = U/2$.\\
$\dot{N}$&$-\frac{\xi N}{\trh'}$ & Rate of change of the total number of bound stars.\\
$\rhdot $&$ \frac{\mu \rh}{\trh'}$ & Rate of change of the half-mass radius.\\
$\dot{\mm} $&$ \frac{\gamma\mm}{\trh'}$ & Rate of change of the mean mass of bound stars.\\
$\dot{\kappa} $&$ \frac{\lambda\kappa}{\trh'}$ & Rate of change of the energy form factor, related to the density profile.\\
$\msd $&$ \ms\frac{\ms_1 - \ms}{\trh'}$ & Rate of change of the concentration parameter for evolving stars. Defined so as to express the efficiency of the adiabatic expansion caused by stellar evolution [i.e. $\dot{E}/|E| = -\ms\dot{\mm}/{\mm}$, and so $\rhdot/\rh = -(\ms-2)\dot{\mm}/{\mm}$].\\
$\mmsed $&$ \frac{\gev\mm}{\trh'}$ & Rate of change of the mean mass of stars due only to the in situ mass-loss caused by stellar evolution. If escaping stars have no preferential mass (e.g. all masses of stars are ejected), $\mm \equiv \mmse$.\\
%\vspace{4mm}
\hline
  \multicolumn{3}{c}{Dimensionless (Differential) Parameters}\\
\hline
$\xi $&$\xie+\xii$ & Dimensionless escape rate per $\trh'$ due to both direct and induced mechanisms.\\
$\xie $&$\F\xi_0(1-\Pa)+\left[f+(1-f)\F\right]\frac35\zeta\Pa$ & Direct relaxation driven escape rate. The first term comes from internal effects \citep{BHH2002} and the second from mass-loss due to a tidal field (Paper I and references therein).\\
$\xii $&$ \find\gev $ & Dynamical escape rate per $\trh'$ induced by stellar evolution.\\
$ \mu $&$\epsilon-2\xi+2\gamma+\lambda$ & Dimensionless change of $\rh$  per $\trh'$ due to dynamical and stellar evolution. Responds so as to maintain the balance of energy (i.e. by conservation of energy).\\
$ \gamma $&$\gev+\gdis$ & Dimensionless change in $\mm$ due to both stellar evolution and the preferential ejection of low-mass stars.\\
$ \gev $&$ -\frac{\nu\trh'}{t}\frac{\mmse}{\mm}$ & Dimensionless change in $\mm$ due to stellar evolution.\\
$ \gdis $&$ \left[1-\frac{\me}{\mm}\right]\Seg\U\xi$ & Dimensionless change in $\mm$ due to the preferential ejection of low-mass stars.\\
$ \lambda $&$ \begin{cases}
0, & n < \nc/2,\\
\left(\kappa_1-\kappa\right)\left(\frac{2n}{\nc}-1\right), & {\rm otherwise},
\end{cases}
 $ & Dimensionless change in the energy form factor, most significant just prior to core collapse (e.g. see Paper II). In Paper II an alternative definition is used to represent the gravothermal catastrophe. However, since SCs with MFs do not undergo the gravothermal catastrophe, we use an approximate form in this study.\\

\hline
\multicolumn{3}{c}{Variable Factors}\\
\hline
$ \Pa $&$ \left(\frac{\rhrj}{\rhrjr}\right)^z\left[\frac{N\log(\gc N_1)}{N_1\log(\gc N)}\right]^{1-x}$ & Function parameterising the rate of escape due to a tidal field (Paper I).\\
$\F $&$ \begin{cases}
0, & n < \nc/2,\\
\left(\frac{2n}{\nc}-1\right), & \nc/2 \leq n \leq \nc,\\
1, & \nc < n,
\end{cases} $ & Smoothing factor to connect the escape rate in unbalanced evolution to the escape rate in balanced evolution. In Paper II it was assumed $\F = \rcrh^{\rm min}/\rcrh$, although $\rc$ is not available for this case. We therefore use an approximation that behaves in an equivalent manner.\\
$\tse(m) $ & $\tse(\mup)\left[1+\frac{\ln (m/\mup)}{ \ln (\mup/\mupi)}\right]^{a} $& Time before the supernova of the most massive star, as a function of the upper limit of the MF ($m_{\rm up}$). Based on the analytic description of \citet{HPT2000}: see Fig.~\ref{f:H-fits}. This function can be inverted to give $\mup(t)$ as a function of $t$.\\
$\Seg $&$ [(\ms-3)/(\ms_1-3)]^q $ & Factor relating the mass segregation to the depletion of low-mass stars.\\
$\U $ & $ (m_{\rm up}(t)-\mm)/(m_{\rm up}(t)) $ & Factor to ensure $\mm < m_{\rm up}(t)$ at all times.\\
$\me $ & $ \X\left(\mm-m_{\rm low}\right)+m_{\rm low} $ & (Average) mass of an escaping star.\\
$\find$ & $\begin{cases}
              \Y\left(\rhrj-\rhrjr\right)^b, &\rhrj > \rhrjr, \\
		0, &{\rm otherwise.}
              \end{cases}$ & Approximation for the relationship between induced escape to mass-loss through stellar evolution; see \citet{LBG2010}.\\
$\psi(t) $&$\begin{cases}
            \pr, &t \le \tse, \\
	    (\pr-\po)\left[\frac{t}{\tse(\mup)}\right]^{y}+\po, &t > \tse.
            \end{cases}
$ & Modification to the standard definition of $\trh$, adjusting to account for the presence of a MF. Approximately represents $\psi = \overline{m^{\beta}}/\overline{m}^{\beta}$: see \citep{SH1971}.\\
\hline
  \multicolumn{3}{c}{Derived cluster properties}\\
\hline
$\trh $&$0.138\frac{\left(N\rh^3\right)^{1/2}}{\sqrt{G\mm}\log(\gc N)}$& Mean relaxation time of stars within the half mass radius \citep{S1987}.\\
$\trh' $&$\frac{\trh}{\psi(t)}$& Modified relaxation time of stars within the half mass radius, adjusted to consider the presence of a MF. See \citep{SH1971}.\\
$n$ & $ \int^{t}_0 \frac{\dr t}{\trh'} $ & Number of modified relaxation times that have elapsed at time $t$. \\
$\rj $&$ \rg\left[\frac{N\mm}{2\mg(<\rg)}\right]^{\frac13}$ & Jacobi (tidal) radius for an isothermal galaxy halo. For a point-mass galaxy the factor of 2 in the denominator is replaced by a 3. The mass contained within the galactocentric (orbital) radius is defined by $\mg(<\rg)$. \\
\end{tabular}
\label{t:parasum}
\end{table*}
\renewcommand{\arraystretch}{1.0}

\renewcommand{\arraystretch}{1.4}
\begin{table*}
\centering
\caption{Definitions of constants used by \textsc{emacss} to define cluster evolution. The values of these constants are taken from literature where possible, or calibrated against $N$-body data when not.}
\begin{tabular}{  r@{\;=\;} m{15cm} }
$\zeta$ & Fractional conduction of energy for clusters with globular cluster like mass functions. In Paper I we find (for clusters of single-mass stars) $\zeta_0 = 0.1$. For this study, we retain this definition although note that the conduction of energy is time dependent (owing to our redefinition of $\trh'$).\\
$y$ & Power-law exponent used to express the width of the stellar MF as a function time. \citet{GBHL2010} show that a good approximation to the theory of \citet{SH1971} is provided by $y = -0.3$.\\
$\gc$  & Argument of the Coulomb Logarithm. From \citet{GH1996}, $\gc = 0.11$ for single-mass clusters and $\gc = 0.02$ for multi-mass clusters.\\
$\xi_0$  & Dimensionless escape rate of an isolated cluster. From Paper I (for clusters of single-mass stars), $\xi_0 = 0.0141$.\\
$\rhrjr$  & Ratio of $\rh$ to $\rj$. From \citet{H1961} for RV filling clusters of single-mass stars, $\rhrjr = 0.145$. For multi-mass clusters, we measure from Fig.~\ref{f:N-rhrj} a scaling relationship between $\rhrjr$ and $N_1$, and show that $\rhrjr = 0.22$ lies upon the track of all our simulations where $N_1 = 1000$.\\
$z$ & Power-law exponent used to express the scaling of $\xie$ with $\rh$ around $\rhrjr$, see \citet{GB2008}. It is shown in Paper I that $z=1.61$ for clusters of equal-mass stars.\\
$x$ & Power-law exponent used to express the scaling on $\xie$ with $N$ due to the escape time effect. From \citet{B2001}, $x=0.75$.\\
$\X$ & Factor determining the mean (average) mass of ejected stars. If $\X = 0$, $\me = m_{\rm low}$, while if $\X = 1$, $\me = \mm$ and $0 \le \X \le 1$. \\
$N_1$ & Scaling factor, defining an ideal cluster for which the ratio $\rhrj=\rhrjr$. From Paper II (for clusters of single-mass stars), $N_1 \simeq 15000$. For multi-mass clusters, we measure from Fig.~\ref{f:N-rhrj} a scaling relationship between $\rhrjr$ and $N_1$, and show that $N_1 = 1000$ lies upon the track of all our simulations where $\rhrjr = 0.22$.\\
$\nu$ & Power-law exponent used to express the change of $\mm$ with respect to time due to stellar evolution (typically) at the high mass end of the MF. From \citet{GBHL2010} for a \citet{K2001} initial mass function (IMF), $\nu = 0.07$.\\
$\tse$ & Time before the start of stellar evolution (i.e. main sequence lifetime of the most massive star present). From stellar evolution models with a maximum star mass of 100 $M_{\odot}$, $\tse=3.3\;{\rm Myr}$. \\
$a$ & Power defining the relationship between main sequence lifetime and stellar mass. Fit in Fig.~\ref{f:H-fits} against the stellar evolution theory of \citet{HPT2000}, where it is shown $a = -2.7$ provides a satisfactory fit. \\
$\nc$ & Number of modified relaxation times ($\trh'$) that elapse prior to the start of balanced evolution. We also define balanced evolution as beginning at time $\tdyn$, although this cannot be calculated at $t=0$. \\
$\msi$ & Scaling factor defining the efficiency of energy production in a fully mass segregated cluster.\\
$f$ & Factor by which the escape rate is reduced in unbalanced evolution. From Paper II, $f = 0.3$.\\
$\kappa_0$ & Initial energy form factor, $\kappa_0 \simeq 0.2$ for a \citet{K1966} or \citet{P1911} model.\\
$\kappa_1$& Energy form factor during balanced evolution. From Paper II, $\kappa_1 \simeq 0.24$.\\
$q$ & Scaling index that relates the degree of mass segregation to the degree of low-mass depletion. \\
$\Y$ & Factor defining the $\rhrj$ dependence of $\xii$. \\
$b$ & Power scaling the induced mass loss to RV filling factor, determined by comparison against $N$-body data.\\
\end{tabular}
\label{t:parascal}
\end{table*}
\renewcommand{\arraystretch}{1.0}

There are several new parameters for which literature values are not available. Specifically, for a cluster in isolation the maximum efficiency factor for energy production by stellar evolution, $\msi$, is not known, and the number of scaled $\trh'$ that pass before balanced evolution begins ($\nc$) is not defined. Furthermore, we use $N$-body data to determine an approximate value for the excess mass of a typical ejected star ($\X$), and the index defining the degree to which mass segregation affects low-mass depletion ($q$).

\begin{figure}
\centering
\includegraphics[height=80mm]{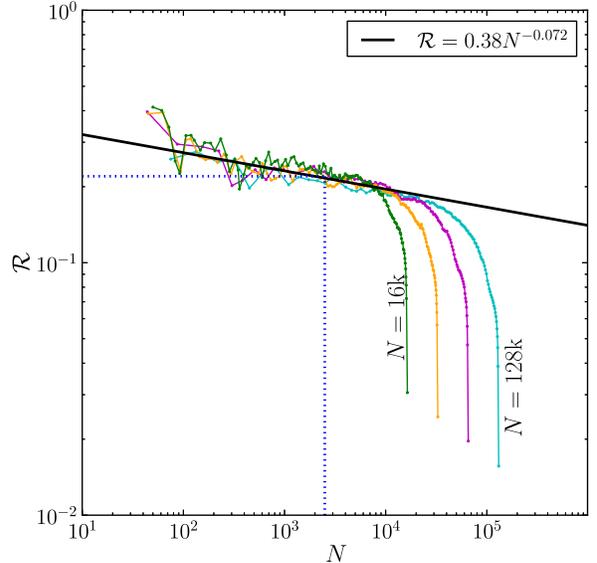}
\caption{Evolution of $\rhrj$ as a function of $N$ for $N$-body simulations. In this figure, the evolution of time goes from right to left. The simulations have $N$=16, 32, 64 and 128k stars as labelled, and begin with $\rh = 1$ pc. All the clusters begin their evolution with expansion, but eventually evolve toward the same track in the $N-\rhrj$ space once RV filling. The (black) solid line is a single-power law fit, while the (blue) dotted lines demonstrate a reference pair of $N$ and $\rhrj$ that lie upon the track of common evolution, well within the region in which all clusters are RV filling. We let this pair be $N_1$ and $\rhrjr$, and therefore choose $N_1 = 1000$ and $\rhrjr = 0.22$ for multi-mass clusters with stellar evolution.}
\label{f:N-rhrj}
\end{figure}

Because the direct escape rate from multi-mass clusters is not necessarily the same as for clusters of single-mass stars, we need to redefine $N_1$, $z$ and $\rhrjr$. The definitions of $N_1$ and $\rhrjr$ are highly (though not totally) degenerate (see Paper I appendix A), and define coordinates in an $(N,\rhrj)$ space through which (contracting and RV filling) clusters will eventually evolve. The relationship between these parameters is illustrated in Fig.~\ref{f:N-rhrj}, from which we choose a pair of parameters $(N_1, \rhrjr) = (1000,0.22)$ for scaling. We have not used the original definition $\rhrjr = 0.145$ from \citet[][]{H1961}, Paper I and II since this value is significantly lower than the $\rhrj$ values measured in any of our new $N$-body simulations. While it would be possible to extrapolate to a much higher $N_1$ where $\rhrj=0.145$, the corresponding $N_1$ is well above our range of $N$ under test and would significantly change $\xie$ due to the $N$ dependence of the Coulomb logarithm. 

Finally, our definition of the efficiency of induced mass loss is new and therefore has not been previously measured. To this end, we define values for $\Y$ and $b$ by comparison against $N$-body simulations of RV filling clusters. The parameters that are optimised by comparison against $N$-body data in section~\ref{s:vali} are $\msi$, $\nc$, $z$, $\X$, $\Y$, $q$ and $b$.

\subsection{The model (\textsc{emacss})}

Using the quantities, properties, and equations defined in tables~\ref{t:epsilon},~\ref{t:parasum} and~\ref{t:parascal}, we extend the theoretical framework upon which \textsc{emacss} is based. As in Papers I and II, we solve the various differential equations for incremental time-steps using Runge-Kutta numerical integration.

In Paper I, it is trivial to calculate an appropriate time-step size for the Runge-Kutta scheme, since clusters of single-mass stars have only one timescale (relaxation) defining their evolution. Hence, the step-size is always set to be $0.1\trh$. In Paper II however, a second timescale (the core relaxation timescale) becomes relevant, and becomes significantly shorter than $0.1\trh$ (and hence paramount for accuracy) for the the runaway stages of core collapse. To compensate for this, in Paper II a combined time step is used including both $\trh$ and the core relaxation time. For multi-mass clusters there are again two distinct timescales: relaxation, and stellar evolution (which has a timescale $\simeq t$; see equation~\ref{eq:label}). A time-step optimised for stellar evolution will be shorter than relaxation when $t$ is small, but will quickly grow to become much larger than $\trh$. 

For this version of \textsc{emacss}, we update our integrator to an adaptive fifth order scheme, in which we use the difference in truncation error between the fourth and fifth order schemes to monitor, adjust and optimise step size \citep[e.g. see][]{F1969}. In addition to this, in order to avoid step size underflow (time steps that approach 0) we assign a lower limit to the time step such that 
\begin{align}
\frac{1}{t_{\rm step}} < \frac{10^6}{\trh}+\frac{10^6}{t},
\label{eq:tstep}
\end{align}
where the first term is relaxation and the second term stellar evolution. 
 
The sequence of calculations performed by \textsc{emacss} is as follows:
\begin{enumerate}
\item{An (estimated) duration of the next time-step is computed from previous time step duration and accuracy.}
\item{The change in energy during the upcoming time-step, $\epsilon$ is computed by the formulae in table~\ref{t:epsilon}, while the dimensionless constants ($\xie, \xii, \mu, \gev, \gdis, \ms_1, \lambda$) and characteristic properties ($\trh, \trh', \rj, \find$) are determined.}
\item{A fifth order Runge-Kutta integration step is applied to equations~(\ref{eq:EdotE}) through~(\ref{eq:mmdotmm}), updating parameters as required. The new state of the cluster (e.g. $N$, $\mm$, $\rh$, $\kappa$) is found.}
\item{The truncation error due to the approximations used by numerical the integrator is examined. If it is large {($>10^{-4}$ per cent)}, the code repeats from step (i) using a smaller time step. If it is acceptable, or if further reduction would cause an overly small time step {(see equation~\ref{eq:tstep})}, the step is completed and the code progresses. {The values are chosen such that a cluster's lifecycle is described by $\simeq 1000$ time-steps, which is fewer than previous versions on account of the efficiency increase offered by the adaptive Runge Kutta scheme.} }
\item{Specified values are output.}
\item{Steps (i) through (v) are repeated until {$N \le 200$, which is consistent with papers I and II ($N=200$) and represents the `breakdown' of balanced evolution due to stochastic effects.}}
\end{enumerate}

For the sake of simplicity, the working of \textsc{emacss} is performed in $N$-body units \citep[i.e., $G =N\mm = -4E = \rv = 1$][]{HM1986}, although conversion to physical units (Myr, $M_{\odot}$, ${\rm pc}$) is trivial. In order to facilitate easy comparison to `real' SCs, this conversion is (by default) performed at output.

\section{General Results}
\label{s:vali}

As \textsc{emacss} is intended to model SCs evolving with a range of initial conditions and tidal environments, we benchmark the efficacy of our prescription against the $N$-body database outlined in section~\ref{s:nb}. We first optimise our prescription for simple (isolated) models, before increasing the complexity through the addition of a tidal field. For isolated clusters we have just two undefined parameters ($\msi$ and $\tdyn$), while there are five additional parameters for tidally limited clusters ($z$, $\X$, $\Y$, $b$ and $q$). We first fit $\msi$ and $\tdyn$ for our simulations of isolated clusters in section~\ref{s:iso}, before fitting $z$, $\X$, $\Y$, $b$ and $q$ for our tidally limited simulations in section~\ref{s:tidal}. We also show some additional comparisons against clusters in additional tidal fields and on eccentric orbits in appendix~\ref{ap:extras}. Throughout, we use $(N_1, \rhrjr) = (1000,0.22)$ as demonstrated in section~\ref{s:all}.

\subsection{Isolated clusters}
\label{s:iso}

\begin{figure}
\centering
\includegraphics[width=80mm]{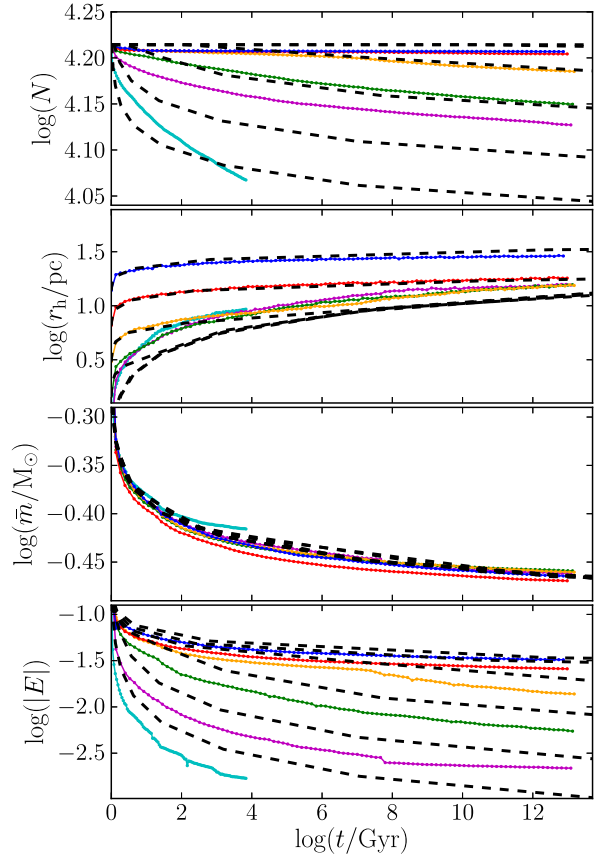}
\caption{Comparison of $N$ (top), $\rh$, $\mm$, and $E$ (bottom) as a function of $t$ for the model described in sections~\ref{s:sources}, \ref{s:sinks1} and~\ref{s:sinks2} to $N$-body data for isolated clusters from section~\ref{s:nb}. The simulations are chosen such that the range of $\trhi$ is logarithmically spaced between 0.03 and 3 Gyr, with the topmost (blue) track being that of the $\trhi = 3$ Gyr simulation and the lowest (cyan) being the $\trhi = 0.03$ Gyr simulation.}
\label{f:iso-fit}
\end{figure}

Fig.~\ref{f:iso-fit} shows the evolution of several parameters ($N$, $\rh$, $\mm$, $E$) from $N$-body simulations of isolated SCs (see section~\ref{s:nb}). Over-plotted are the equivalent tracks predicted by \textsc{emacss} for clusters with the same initial conditions, with our best fitting parameters of $\msi = 4$ and $\nc = 12.5$.

The evolution of $\mm$ in the third panel is well described by a single-power law for isolated clusters, using the same $\nu = 0.07$ as measured by previous studies \citep[e.g.][]{GBHL2010}. This is consistent with the loss of mass due to stellar evolution, which is (for isolated clusters) the predominant cause of mass loss. Further to this, an expansion of $\rh$ such that $\rhdot/\rh = -(\ms-2)\mmsed/\mm$ is shown to provide a adequate description of the $\rh$ evolution, implying that $\ms$ evolves in a similar manner to that suggested in section~\ref{s:sinks1}, and that energy evolves such that $\dot{E}/|E| = -\ms\mmsed/\mm$. The start of balanced evolution is evident as an upturn in the $\rh$ evolution in both $N$-body data and the predictions of \textsc{emacss}, suggesting that our value $\nc = 12.5$ is appropriate for these clusters. 

We find that the escape rate from isolated clusters with stellar evolution is somewhat over-predicted if we retain the description of escape from Paper I (see equation~\ref{eq:xi}), whereby (in isolated clusters) the fraction of escapers $\xi_0 = 0.0142$ per $\trh'$. The over-prediction of $\xi$ is accounted for by our use of $\trh'$ which (at early times) is a factor $\sim 8$ smaller than $\trh$. We find this over-prediction is reduced if we redefine $\xi_0 = 0.0075$. {The value $\msi = 4$ chosen is somewhat lower than expected. However, there is some data (specifically, the $\rh$ increase of the short initial $\trh$ clusters) to suggest that the mass segregation is higher than $\msi = 4$ within the first $\simeq$ 1 Gyr, which is evidenced by the under-predicted expansion and energy increase produced by \textsc{emacss}. A higher $\msi$ results in excessive expansion of long $\trh$ clusters at times $\gtrsim 10$ Gyr. This suggests that the mass segregation is more severe than $\msi = 4$ when high mass stars are still present, although the effect reduces as high mass stars evolve. The value obtained, $\msi = 4$, therefore represents a compromise value.} Despite this, $\rh$ is reproduced to within a factor of two, suggesting a good qualitative description for the evolution of isolated clusters is obtained by our prescription, and note anyway that the numerical discrepancy is acceptable since realistic clusters do not evolve in a completely isolated state.

\subsection{Clusters in tidal fields}
\label{s:tidal}

\begin{figure*}
\centering
\includegraphics[width=160mm]{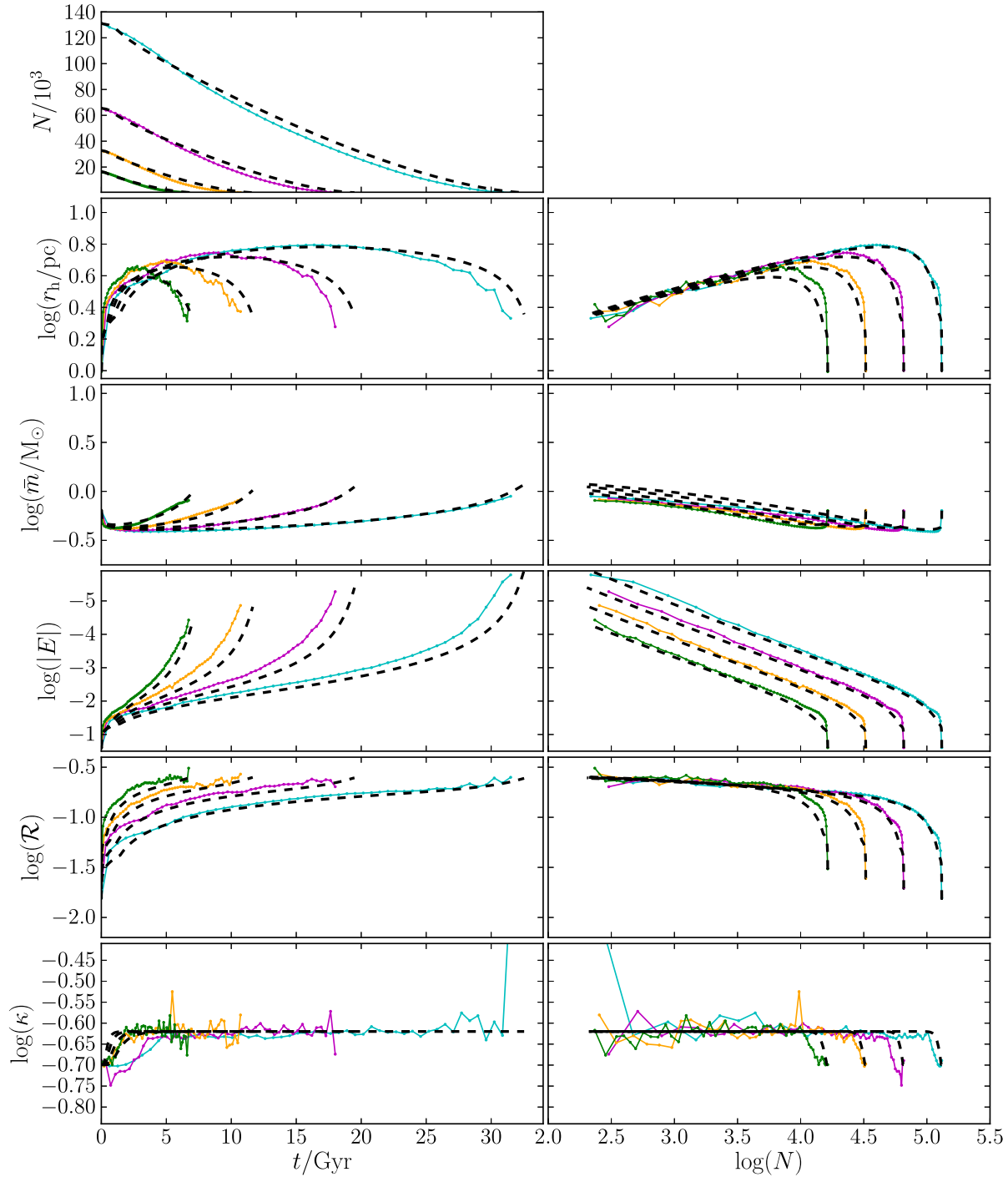}
\caption{Comparison of the \textsc{emacss} model described in sections~\ref{s:sources}, \ref{s:sinks1} and~\ref{s:sinks2} to $N$-body data for a series of $N$-body simulations of clusters in tidal fields (approximated as isothermal halos with $\vg = 220$ km${\rm s}^{-1}$ and $\rg = 8.5$ kpc). The left column plots $N$, $\rh$, $\mm$, $E$, $\rhrj$ and $\kappa$ as a function of time, while the right column plots the same properties as a function of $\log(N)$ .The initial conditions of the $N$-body simulations are described in section \ref{s:nb}, with $N_0$=128k (cyan track; longest surviving), $N_0$=64k (magenta), $N_0$=32k (orange), $N_0$=16k (green; shortest surviving) stars. The clusters are allowed to evolve until their eventual dissolution. For these SCs, $\rhi = 1$ pc.}
\label{f:tfit1}
\end{figure*}

\begin{figure*}
\centering
\includegraphics[width=160mm]{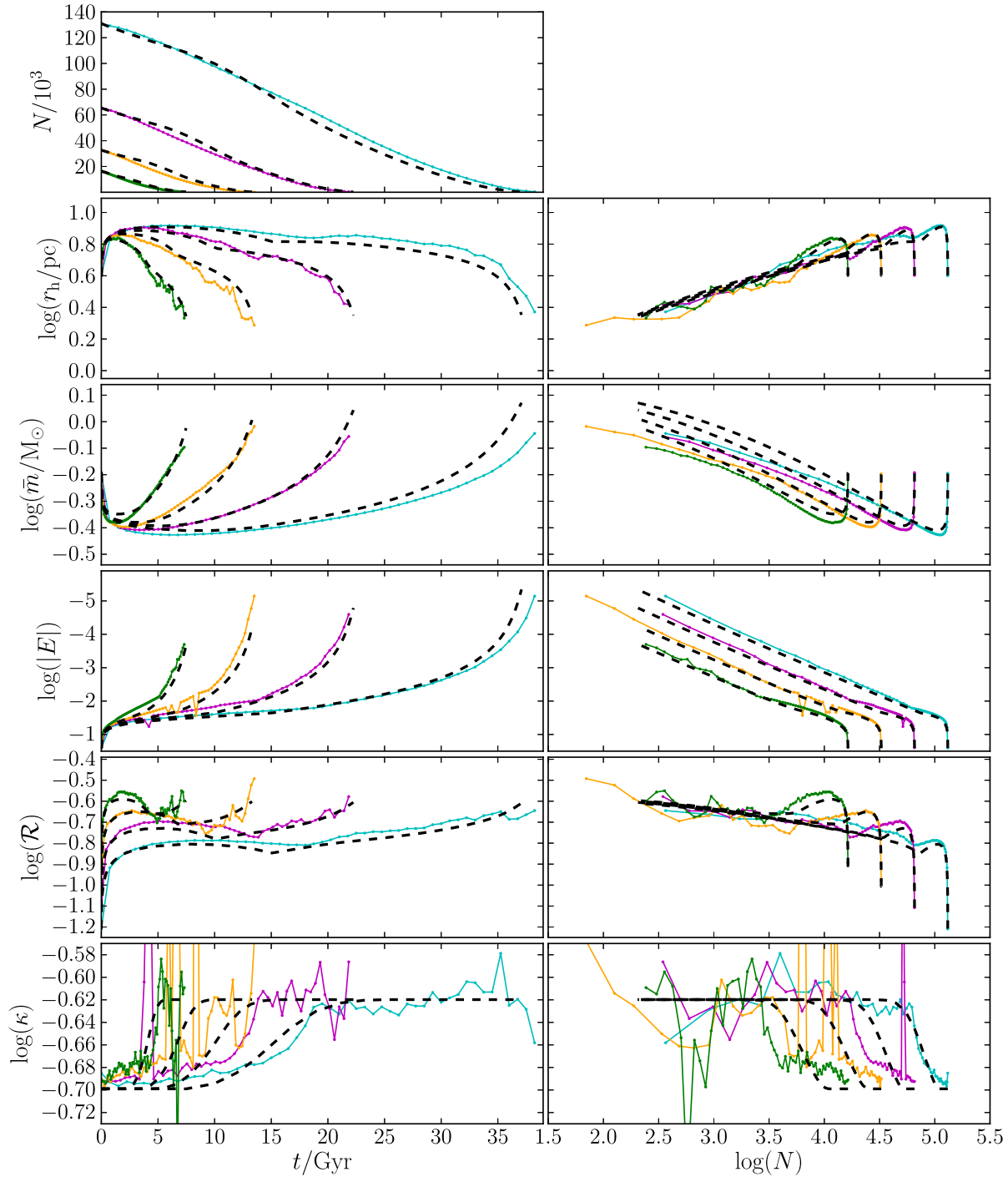}
\caption{As Fig.~\ref{f:tfit1}, but with simulated clusters of larger initial radii. For these clusters, $\rhi = 4$ pc.}
\label{f:tfit4}
\end{figure*}

\begin{figure*}
\centering
\includegraphics[width=160mm]{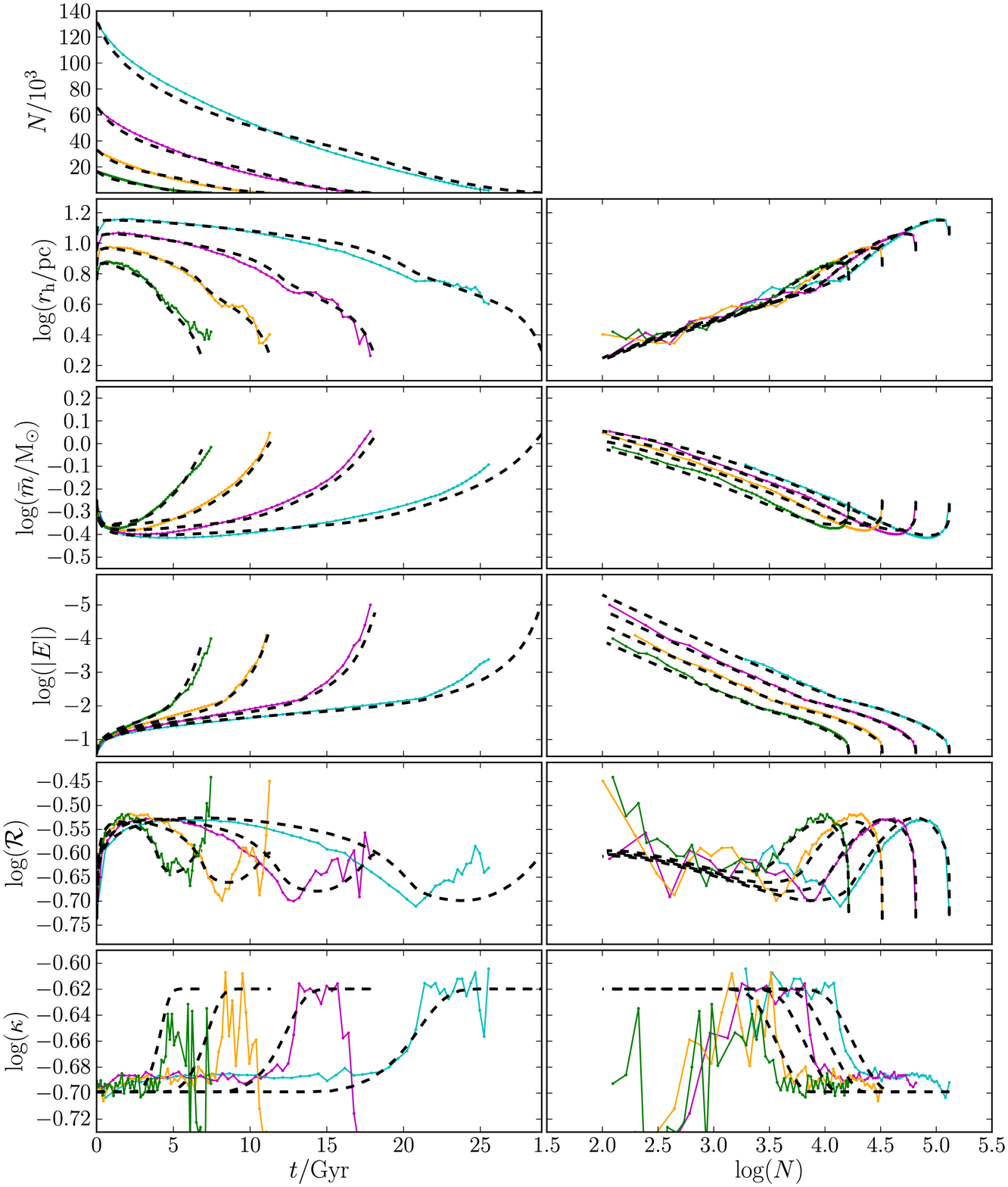}
\caption{As Fig.~\ref{f:tfit1}, but with simulated clusters that are initially RV filling. For these clusters, $\rhrjo \simeq 0.19$. As the MF for the comparison $N$-body simulations has lower $m_{\rm up}$, $\tse$ is slightly longer (see equation~\ref{eq:tse}) and $\mm_0 = 0.547$ in both the $N$-body data and \textsc{emacss}.}
\label{f:tfitRF}
\end{figure*}

Figs.~\ref{f:tfit1},~\ref{f:tfit4} and~\ref{f:tfitRF} show the evolution of all parameters for the $N$-body simulations of SCs in tidal fields at $\rg = 8.5$ kpc. Our remaining simulations -- those of initially RV filling clusters at $\rg = 2.8$ kpc or $\rg = 15$ kpc, and those on eccentric orbits -- are additionally shown in appendix~\ref{ap:extras}. Once again, in all cases we have over-plotted the results of \textsc{emacss} using the best fitting $z$, $\X$, $\Y$, $b$ and $q$, while we use the $\ms=4$ and $\nc=12.5$ defined in section~\ref{s:iso}. {For these clusters, \textsc{emacss} provides} a reasonable estimation of the properties of our $N$-body simulations, and exhibits similar (qualitative) features throughout.

We find optimal values for our parameters to be $z = 2.0$, $\X = 0.55$, $\Y=90$, $b = 1.35$ and $q = 2$. In this case, $z$ is somewhat higher than for the case of single-mass mass clusters, which may be caused by the mass segregation of the different stellar species. Because of mass segregation, low-mass species will form the majority of the SC halo, which (for the same density profile) will lead to increased numbers of stars in the outermost orbits. Consequently, the escape rate $\xie$ is more sensitive to $\rhrj$, as suggested by a higher value of $z$. This theory is also supported by the value of $\X = 0.55$, since this implies that the escaping stars are (on average) mid-way between $\mm$ and $m_{\rm low}$, and are therefore have $m < \mm$. If the majority of escapers are assumed to originate in the cluster halo, the value $\X < 1$ confirms that the halo consists mainly of low mass stars.

The power-law scaling for the rate of low-mass star depletion, $q$, is a free parameter, for which we find $q=2$ fits well. This implies that the effects of mass segregation (represented by $\ms$) are felt somewhat more strongly amongst the low mass species (segregating outward into the halo) of the halo than the high mass species (segregating inward toward the cluster core). If $q$ is varied, we find that $q \ll 2$ represents too weak a relationship between mass segregation and low mass depletion (and too late an upturn in $\mm$), while {$q \gg 2$} is too strong (with the upturn of $\mm$ occurring too early). The value $q=2$ implies that mass segregation is slightly more strongly felt by low mass species, perhaps as a consequence of high mass stars starting to fall inward early and leaving a low-mass halo behind.

Since $\X$ controls the `typical' escaper mass and not (directly) a physical value, the value of $\X$ is (reasonably) invariant throughout cluster evolution. This means that the mass of a (typical) escaping star will (correctly) increase as $\mm$ grows. This increase is also shown in the $N$-body data, and slows appropriately once $\mm$ approaches $\mup(t)$ (due to the increasing $\U$ parameter). For the sake of simplicity, we retain a time invariant $\X$, and encompass the variation of $\me$ with respect to $\mm$ through $\U$. {Further, we note that evolving stars are ipso facto assumed to loose all their mass, and hence that $p = 0$ (section~\ref{s:ms}) throughout cluster evolution. Both these factors could be incorporated into a more exact model, and allowed to vary so as to offer an improved description of the evolution of cluster stars. A manner in which full variation of the mass function could be built} upon the work of \citet{LBG2013}, although for the sake of brevity we do not attempt this here, {and caution that such models will be liable to significant stochastic effects.}

Finally, the values found for $\Y=90$ and $b=1.35$ suggests that the induced mass loss increases quite rapidly once $\rhrj > \rhrjr$. This increased mass loss would imply that substantially overfilling clusters are unlikely to survive for a significant period of time. Instead, such RV overflowing clusters will lose their outermost stars very quickly, with the result that $\rh$ shrinks faster than $\rj$ and the cluster returns to a filling- or under-filling state.

As in isolated clusters, the early unbalanced evolution of under-filling clusters is dominated by expansion, caused by the energy increase due to stellar evolution's mass loss. If SCs expand sufficiently to become RV filling before the end of the unbalanced phase, the rate of escape due to relaxation increases and the cluster will contract owing to a decreasing $\rj$. At the moment of core collapse however, $\epsilon$ instantaneously increases by a factor $\simeq 2-3$, potentially leading to a secondary expansion (and later contraction) during balanced evolution. We find therefore that many of our RV under-filling clusters experience `double peaked' radius evolution, depending on the range of the MF and the ratio $\mathcal{R}$ at the time of core collapse. 

For RV filling clusters, unbalanced evolution once again begins with an adiabatic expansion. However, in this case, the expansion very quickly causes the cluster to become over-filling, and hence always results in significant induced mass-loss, meaning that the majority of the stars of a filling SC escape before core collapse. Furthermore, once a filling cluster reaches core collapse, it remains RV filling and hence continues to contract for the remainder of its life-cycle. For both filling- and under-filling SCs, the evolution of $N$ is a smooth function of $t$ since both $\xie$ and $\xii$ are functions of $\rhrj$, despite $\epsilon$ varying discontinuously.

{Despite general success, there are several minor discrepancies between the $N$-body data and the \textsc{emacss} models for the various simulations. Some of these are random effects occurring in the $N$-body simulations (such as the fluctuations of post-collapse $\rh$ seen in the $N$-body simulations of Fig.~\ref{f:tfitRF} or the energy evolution of the 64k run in Fig.~\ref{f:tfit4}, which are simply the results of stochastic noise in the simulations). Others are accounted for by the various approximations made by our code -- equation~(\ref{eq:xi}) approximates the geometry of the Jacobi surface and induced mass loss, which leads to the majority of the minor oscillation of $N$ (or $\rhrj$, since $\rj \propto N^{1/3}$) in \textsc{emacss} when compared to $N$-body data. We also find that the $\rhrj$ evolution shown in Fig.~\ref{f:tfitRF} is a somewhat less smooth curve for \textsc{emacss} than for the $N$-body data, which is most likely a consequence of the simplifications in the approximation for induced mass-loss. Induced mass-loss is likely to be a more complex process than appreciated by equations~(\ref{eq:ind}) and~(\ref{eq:find}), as suggested in \citet{LBG2010}. Our most approximate overall quantity is the fitting function used for $\kappa$, which demonstrates the loosest fit to $N$-body data and perhaps exemplifies the weakest area of our code. This quantity is, however, of negligible importance for the overall properties we seek to model and which can be compared to observations.}

The mean mass of stars in tidally limited $N$-body simulations shows an initial power-law decrease similar to that of isolated clusters. However, as predicted in section~\ref{s:dmm}, this eventually turns into an increasing $\mm$ due to the preferential ejection of low-mass stars. We find that \textsc{emacss} recovers this upturn to occur correctly for the majority of SC's lifetime. 

\section{Conclusions}
\label{s:conc}
We have enhanced the \textsc{emacss} prescription for star cluster (SC) evolution first presented in \citet{AG2012,GALB2014} and originally based upon the work of \citet{H1961}, \citet{H1965}, and \citet{GHZ2011}. Our enhancements incorporate the results of \citet{LBG2010} (to describe the effect of stellar evolution upon escape) and \citet{GBHL2010} (to describe the effect of stellar evolution upon half-mass radius), and include the addition of a full stellar initial mass function, stellar evolution, mass segregation, and an improved model of the pre core-collapse (`unbalanced') evolution of SCs. We have compared this prescription against a series of $N$-body simulations, and demonstrate that it remains accurate over the entirety of a SC's lifetime throughout a range of tidal environments.

We find three (almost independent) energy sources driving various epochs of SC evolution. For the first $\lesssim 3$~Myr, the change in energy is $\simeq 0$, as little stellar- or dynamical evolution occurs in this interval. Furthermore, binaries are unlikely to be forming except for in very compact clusters\footnote{unless primordial binaries are present, e.g. \citep{THH2006}}. \citep{SOC1987}.  After a period of time that depends upon the mass of the highest mass stars present ($\tse \simeq 3.3$~Myr for stars of $100\,M_{\odot}$), the highest mass stars will begin to evolve off the main sequence, leading into a second stage of evolution with energy `produced' by stars evolving within the SC potential. This is an `unbalanced' phase, where energy production is controlled by the rate of stellar evolution and is not in balance with two-body relaxation. The final stage is a balanced phase, similar to that discussed in \citet{GHZ2011}. The unbalanced stage can be very short for very compact clusters with $\trh \simeq 1$ Myr, which can undergo core collapse during the first few Myr; for these clusters, binaries are formed early, and the cluster enters balanced evolution shortly after formation.

SCs respond to the input of energy through stellar evolution mainly by expansion and the escape of stars. If SCs are initially Roche volume (RV) under-filling, escape is at first negligible and a good approximation for the expansion is given by $\rhdot/\rh \propto -\dot{\mm}/\mm$, with the constant of proportionality depending upon mass segregation. As SCs become increasingly RV filling, the rate of induced escape increases, leading to a decreasing rate of expansion. However, even for initially RV filling clusters, early evolution is dominated by expansion, albeit with significant numbers of escapers.

The proportionality between $\dot{\mm}/\mm$ and $\rhdot/\rh$ will vary over the course of unbalanced evolution, since the energy released by stars evolving depends upon the specific potential where stellar evolution is occurring. Mass loss substantially occurs through the winds and supernovae of the most massive stars, such that the energy release will correlate with the location of high mass stars (in particular, their location at the time of their death). By extension, the location of the high mass stars will strongly correlate with the degree of mass segregation in the cluster \citep{KASS2007}. For an initially homologous distribution of stars, mass segregation occurs due to dynamical friction, which acts on (approximately) a relaxation timescale. The efficiency of energy generation will increase at the same rate. From equation~(\ref{eq:nrg-expand}), we express the energy generation by stellar evolution in a cluster by a factor $\ms$, where $\ms \simeq 3$ for a homogeneous distribution of evolving stars and the maximum segregation factor a cluster can achieve is $\ms = 4$. For our isolated clusters, we find some evidence that $\msi$ may be $>4$ at early times (i.e. an even faster loss of energy occurs) when $\trh$ is short. This would suggest that the maximum degree of mass-segregation is dependent on the upper limit of the mass function. However, in the interests of brevity, we find that $\msi = 4$ provides an adequate compromise value overall.

The definition of the escape rate of stars due to two-body relaxation ($\xie$) is similar to its definition for single-mass clusters \citep{AG2012,GALB2014}. However, the scaling values (i.e. the reference $(N_1,\rhrjr)$ pair through which all clusters evolve once RV filling) take different values. Choosing $N_1 = 10^3$, we find a higher $\rhrjr = 0.22$ for multi-mass clusters than for single-mass clusters. The scaling of $\xie$ with $\rhrj$, $z = 2.0$, is also higher than in the case of single-mass clusters, probably on account of mass segregation. This occurs because halo stars are typically of lower than average mass, and are hence on more distant orbits for a given $\rh$ to be measured. Consequently, the higher value of $z$ represents a $\xie$ that is more sensitive to $\rhrj$.

For RV filling- or over-filling clusters, we find an additional escape process - escape induced by stellar evolution, by the process introduced in \citet{LBG2010}. As stars evolve, the cluster will lose overall mass. This mass loss will reduce the radius of the Jacobi surface, which will leave outlying stars outside the Jacobi surface and unbound from the cluster. We use a very simple prescription for the efficiency of induced escape, such that the extent of this process is controlled solely by $\rhrj$. This ratio is used to express the ease through which outlying stars can be removed from the cluster by the shrinking Jacobi radius. The time taken for this induced escape to start is not accounted for by this simple expression \citep[e.g. the delay time in][]{LBG2010}, although we use reduced `efficiency' $\find$ instead. Overall, we find that the relationship between induced escape $\xii$ and RV filling $\rhrj$ increases rapidly once a cluster becomes RV filling or over-filling, and is negligible for under-filling clusters.

In agreement with \citet{GALB2014}, our results suggest that the mass-loss rate for unbalanced evolution is a factor of $\simeq 3$ smaller than than for balanced evolution, which we attribute to different and less efficient channels through which stars escape. Meanwhile, escaping stars are on average less massive than the mean mass (we find an `average' escaper mass approximately mid-way between the low-mass end of the MF $m_{\rm low}$ and $\mm$ best fits our data). This is indicative that the stars escaping do so with a range of masses, \emph{typically} $\me < \mm$. We note however that $\mm$ will initially decrease (as the cluster mass segregates), but will then increase as a result of the depletion of low-mass stars. We find only a very weak relationship between the degree of mass segregation and the low-mass depletion, from which we infer that lower mass species are preferentially ejected even before becoming the dominant constituent of the cluster halo. This would imply that three-body encounters, and the direct ejection of low-mass stars by relaxation, plays a large role in cluster evolution.

Finally, we find that $\simeq 12.5$ modified relaxation times (where $\trh' = \trh/\psi(t)$) pass before the cluster reaches balance at core collapse, which is comparable with the mass segregation time for multi-mass clusters, and agrees with the time taken for clusters to reach a roughly equilibrium distribution of mass species \citep[e.g. the mass segregation times measured by][]{PZM2002}.

The new version of \textsc{emacss} is publicly available\footnote{https://github.com/emacss/emacss}. The code is bench marked against $N$-body simulations, and can be used to efficiently model several properties of clusters \citep[e.g. for the purpose of statistical studies of synthetic SC populations,][]{SKYK2013,AG2013}. In addition, due to the speed of the prescription, potential applications exist as an iterative tool to search for the initial conditions of observed clusters.

Despite these successes, the code still retains only a descriptions of a few physical observables of clusters (e.g. only the total mass, mean stellar mass, and radius are included, while the mass function, density profile and velocity dispersion profile are not modelled directly). {Moreover, some physical effects: the retention of black holes and neutron stars; and realistic ejection mechanisms for low mass stars are only approximately considered. Finally,} \textsc{emacss} does not directly model eccentric cluster orbits (instead relying on approximation), and does not yet consider changing orbits (e.g. dynamical friction). Further work is therefore forthcoming to improve the description of the tidal field and number of observable SC properties considered. However, the present code produces comparable results to $N$-body simulations for a wide range of initial conditions, and can readily be used to effectively model galactic SCs, using appropriate approximations where required \citep[e.g., see][]{HG2008}. Several studies are forthcoming to explore the use of this code as a population modelling tool (e.g. Pijloo et al. in prep., Alexander et al. in prep.).

\section*{Acknowledgements}
All the authors are grateful to the Royal Society for an International Exchange Scheme Grant between the University of Queensland, Australia, and the University of Surrey, UK, through which this study was made possible. PA acknowledges the UK Science and Technology Facilities Council for financial support via a graduate studentship. MG thanks the Royal Society for financial support via a University Research Fellowship and an equipment grant. H.B. acknowledges support from the Australian Research Council through Future Fellowship grant FT0991052. {Finally, the authors would like to thank our anonymous reviewer for helpful comments and suggestions.}

\bibliography{emacss}

\begin{thebibliography}{}

\bibitem[\protect\citeauthoryear{{Aarseth}}{{Aarseth}}{1999}]{A1999}
{Aarseth} S.~J.,  1999, \pasp, 111, 1333

\bibitem[\protect\citeauthoryear{{Alexander} \& {Gieles}}{{Alexander} \&
  {Gieles}}{2012}]{AG2012}
{Alexander} P.~E.~R.,  {Gieles} M.,  2012, \mnras, 422, 3415

\bibitem[\protect\citeauthoryear{{Alexander} \& {Gieles}}{{Alexander} \&
  {Gieles}}{2013}]{AG2013}
{Alexander} P.~E.~R.,  {Gieles} M.,  2013, \mnras, 432, L1

\bibitem[\protect\citeauthoryear{{Ambartsumian}
  V.}{{Ambartsumian}}{1938}]{A1938}
{Ambartsumian} V. A.,  1938, Ann. Len. State Univ., 22, 19

\bibitem[\protect\citeauthoryear{{Baumgardt}}{{Baumgardt}}{2001}]{B2001}
{Baumgardt} H.,  2001, \mnras, 325, 1323

\bibitem[\protect\citeauthoryear{{Baumgardt}, {Hut} \& {Heggie}}{{Baumgardt}
  et~al.}{2002}]{BHH2002}
{Baumgardt} H.,  {Hut} P.,    {Heggie} D.~C.,  2002, \mnras, 336, 1069

\bibitem[\protect\citeauthoryear{{Baumgardt} \& {Makino}}{{Baumgardt} \&
  {Makino}}{2003}]{BM2003}
{Baumgardt} H.,  {Makino} J.,  2003, \mnras, 340, 227

\bibitem[\protect\citeauthoryear{{Bettwieser} \& {Inagaki}}{{Bettwieser} \&
  {Inagaki}}{1985}]{BI1985}
{Bettwieser} E.,  {Inagaki} S.,  1985, \mnras, 213, 473

\bibitem[\protect\citeauthoryear{{Breen} \& {Heggie}}{{Breen} \&
  {Heggie}}{2013}]{BH2013}
{Breen} P.~G.,  {Heggie} D.~C.,  2013, \mnras, 436, 584

\bibitem[\protect\citeauthoryear{{Chernoff} \& {Weinberg}}{{Chernoff} \&
  {Weinberg}}{1990}]{CW1990}
{Chernoff} D.~F.,  {Weinberg} M.~D.,  1990, \apj, 351, 121

\bibitem[\protect\citeauthoryear{{Ernst} \& {Just}}{{Ernst} \&
  {Just}}{2013}]{EJ2012}
{Ernst} A.,  {Just} A.,  2013, \mnras, p.~554

\bibitem[\protect\citeauthoryear{Fehlberg}{Fehlberg}{1969}]{F1969}
Fehlberg E.,  1969, 4, 93

\bibitem[\protect\citeauthoryear{{Fujii} \& {Portegies Zwart}}{{Fujii} \&
  {Portegies Zwart}}{2013}]{FPZ2013}
{Fujii} M.~S.,  {Portegies Zwart} S.,  2013, ArXiv e-prints

\bibitem[\protect\citeauthoryear{{Fukushige} \& {Heggie}}{{Fukushige} \&
  {Heggie}}{1995}]{FH1995}
{Fukushige} T.,  {Heggie} D.~C.,  1995, \mnras, 276, 206

\bibitem[\protect\citeauthoryear{{Fukushige} \& {Heggie}}{{Fukushige} \&
  {Heggie}}{2000}]{FH2000}
{Fukushige} T.,  {Heggie} D.~C.,  2000, \mnras, 318, 753

\bibitem[\protect\citeauthoryear{{Gieles}, {Alexander}, {Lamers} \&
  {Baumgardt}}{{Gieles} et~al.}{2014}]{GALB2014}
{Gieles} M.,  {Alexander} P.~E.~R.,  {Lamers} H.~J.~G.~L.~M.,    {Baumgardt}
  H.,  2014, \mnras, 437, 916

\bibitem[\protect\citeauthoryear{{Gieles} \& {Baumgardt}}{{Gieles} \&
  {Baumgardt}}{2008}]{GB2008}
{Gieles} M.,  {Baumgardt} H.,  2008, \mnras, 389, L28

\bibitem[\protect\citeauthoryear{{Gieles}, {Baumgardt}, {Heggie} \&
  {Lamers}}{{Gieles} et~al.}{2010}]{GBHL2010}
{Gieles} M.,  {Baumgardt} H.,  {Heggie} D.~C.,    {Lamers} H.~J.~G.~L.~M.,
  2010, \mnras, 408, L16

\bibitem[\protect\citeauthoryear{{Gieles}, {Heggie} \& {Zhao}}{{Gieles}
  et~al.}{2011}]{GHZ2011}
{Gieles} M.,  {Heggie} D.~C.,    {Zhao} H.,  2011, \mnras, 413, 2509

\bibitem[\protect\citeauthoryear{{Giersz} \& {Heggie}}{{Giersz} \&
  {Heggie}}{1996}]{GH1996}
{Giersz} M.,  {Heggie} D.~C.,  1996, \mnras, 279, 1037

\bibitem[\protect\citeauthoryear{{Giersz} \& {Heggie}}{{Giersz} \&
  {Heggie}}{1997}]{GH1997}
{Giersz} M.,  {Heggie} D.~C.,  1997, \mnras, 286, 709

\bibitem[\protect\citeauthoryear{{Giersz} \& {Heggie}}{{Giersz} \&
  {Heggie}}{2011}]{GH2011}
{Giersz} M.,  {Heggie} D.~C.,  2011, \mnras, 410, 2698

\bibitem[\protect\citeauthoryear{{Gnedin} \& {Ostriker}}{{Gnedin} \&
  {Ostriker}}{1997}]{GO1997}
{Gnedin} O.~Y.,  {Ostriker} J.~P.,  1997, \apj, 474, 223

\bibitem[\protect\citeauthoryear{{Harris}}{{Harris}}{1996}]{H1996}
{Harris} W.~E.,  1996, \aj, 112, 1487

\bibitem[\protect\citeauthoryear{{Heggie}}{{Heggie}}{1975}]{H1975}
{Heggie} D.~C.,  1975, \mnras, 173, 729

\bibitem[\protect\citeauthoryear{{Heggie} \& {Giersz}}{{Heggie} \&
  {Giersz}}{2008}]{HG2008}
{Heggie} D.~C.,  {Giersz} M.,  2008, \mnras, 389, 1858

\bibitem[\protect\citeauthoryear{{Heggie} \& {Mathieu}}{{Heggie} \&
  {Mathieu}}{1986}]{HM1986}
{Heggie} D.~C.,  {Mathieu} R.~D.,  1986, in {P.~Hut \& S.~L.~W.~McMillan} ed.,
  The Use of Supercomputers in Stellar Dynamics Vol.~267 of Lecture Notes in
  Physics, Berlin Springer Verlag, {Standardised Units and Time Scales}.
pp 233--+

\bibitem[\protect\citeauthoryear{{Heggie}, {Trenti} \& {Hut}}{{Heggie}
  et~al.}{2006}]{THH2006}
{Heggie} D.~C.,  {Trenti} M.,    {Hut} P.,  2006, \mnras, 368, 677

\bibitem[\protect\citeauthoryear{{H{\'e}non}}{{H{\'e}non}}{1961}]{H1961}
{H{\'e}non} M.,  1961, Annales d'Astrophysique, 24, 369

\bibitem[\protect\citeauthoryear{{H{\'e}non}}{{H{\'e}non}}{1965}]{H1965}
{H{\'e}non} M.,  1965, Annales d'Astrophysique, 28, 62

\bibitem[\protect\citeauthoryear{{H\'{e}non}}{{H\'{e}non}}{1969}]{H1969}
{H\'{e}non} M.,  1969, \aap, 2, 151

\bibitem[\protect\citeauthoryear{{Hills}}{{Hills}}{1980}]{H1980}
{Hills} J.~G.,  1980, \apj, 235, 986

\bibitem[\protect\citeauthoryear{{Hurley}}{{Hurley}}{2007}]{H2007}
{Hurley} J.~R.,  2007, \mnras, 379, 93

\bibitem[\protect\citeauthoryear{{Hurley}, {Pols} \& {Tout}}{{Hurley}
  et~al.}{2000}]{HPT2000}
{Hurley} J.~R.,  {Pols} O.~R.,    {Tout} C.~A.,  2000, \mnras, 315, 543

\bibitem[\protect\citeauthoryear{Hurley \& Shara}{Hurley \&
  Shara}{2012}]{HS2012}
Hurley J.~R.,  Shara M.~M.,  2012, \mnras, 425, 2872

\bibitem[\protect\citeauthoryear{{Hurley}, {Tout} \& {Pols}}{{Hurley}
  et~al.}{2002}]{HTP2002}
{Hurley} J.~R.,  {Tout} C.~A.,    {Pols} O.~R.,  2002, \mnras, 329, 897

\bibitem[\protect\citeauthoryear{{Khalisi}, {Amaro-Seoane} \&
  {Spurzem}}{{Khalisi} et~al.}{2007}]{KASS2007}
{Khalisi} E.,  {Amaro-Seoane} P.,    {Spurzem} R.,  2007, \mnras, 374, 703

\bibitem[\protect\citeauthoryear{{King}}{{King}}{1958}]{K1958}
{King} I.,  1958, \aj, 63, 114

\bibitem[\protect\citeauthoryear{{King}}{{King}}{1966}]{K1966}
{King} I.~R.,  1966, \aj, 71, 64

\bibitem[\protect\citeauthoryear{{Kroupa}}{{Kroupa}}{2001}]{K2001}
{Kroupa} P.,  2001, \mnras, 322, 231

\bibitem[\protect\citeauthoryear{{Kruijssen}}{{Kruijssen}}{2009}]{K2009}
{Kruijssen} J.~M.~D.,  2009, \aap, 507, 1409

\bibitem[\protect\citeauthoryear{{Lamers}, {Baumgardt} \& {Gieles}}{{Lamers}
  et~al.}{2010}]{LBG2010}
{Lamers} H.~J.~G.~L.~M.,  {Baumgardt} H.,    {Gieles} M.,  2010, \mnras, 409,
  305

\bibitem[\protect\citeauthoryear{{Lamers}, {Baumgardt} \& {Gieles}}{{Lamers}
  et~al.}{2013}]{LBG2013}
{Lamers} H.~J.~G.~L.~M.,  {Baumgardt} H.,    {Gieles} M.,  2013, ArXiv e-prints

\bibitem[\protect\citeauthoryear{{Larson}}{{Larson}}{1970}]{L1970}
{Larson} R.~B.,  1970, \mnras, 147, 323

\bibitem[\protect\citeauthoryear{{L{\"u}tzgendorf}, {Baumgardt} \&
  {Kruijssen}}{{L{\"u}tzgendorf} et~al.}{2013}]{LBK2013}
{L{\"u}tzgendorf} N.,  {Baumgardt} H.,    {Kruijssen} J.~M.~D.,  2013, \aap,
  558, A117

\bibitem[\protect\citeauthoryear{{Lynden-Bell} \& {Eggleton}}{{Lynden-Bell} \&
  {Eggleton}}{1980}]{LBE1980}
{Lynden-Bell} D.,  {Eggleton} P.~P.,  1980, \mnras, 191, 483

\bibitem[\protect\citeauthoryear{{Lynden-Bell} \& {Wood}}{{Lynden-Bell} \&
  {Wood}}{1968}]{LBW1968}
{Lynden-Bell} D.,  {Wood} R.,  1968, \mnras, 138, 495

\bibitem[\protect\citeauthoryear{{Makino} \& {Aarseth}}{{Makino} \&
  {Aarseth}}{1992}]{MA1992}
{Makino} J.,  {Aarseth} S.~J.,  1992, \pasj, 44, 141

\bibitem[\protect\citeauthoryear{{Nitadori} \& {Aarseth}}{{Nitadori} \&
  {Aarseth}}{2012}]{NA2012}
{Nitadori} K.,  {Aarseth} S.~J.,  2012, \mnras, 424, 545

\bibitem[\protect\citeauthoryear{{Plummer}}{{Plummer}}{1911}]{P1911}
{Plummer} H.~C.,  1911, \mnras, 71, 460

\bibitem[\protect\citeauthoryear{{Portegies Zwart} \& {McMillan}}{{Portegies
  Zwart} \& {McMillan}}{2002}]{PZM2002}
{Portegies Zwart} S.~F.,  {McMillan} S.~L.~W.,  2002, \apj, 576, 899

\bibitem[\protect\citeauthoryear{{Portegies Zwart} \& {Rusli}}{{Portegies
  Zwart} \& {Rusli}}{2007}]{PZR2007}
{Portegies Zwart} S.~F.,  {Rusli} S.~P.,  2007, \mnras, 374, 931

\bibitem[\protect\citeauthoryear{{Renaud}, {Gieles} \& {Boily}}{{Renaud}
  et~al.}{2011}]{RGB2011}
{Renaud} F.,  {Gieles} M.,    {Boily} C.~M.,  2011, \mnras, 418, 759

\bibitem[\protect\citeauthoryear{{Shin}, {Kim}, {Yoon} \& {Kim}}{{Shin}
  et~al.}{2013}]{SKYK2013}
{Shin} J.,  {Kim} S.~S.,  {Yoon} S.-J.,    {Kim} J.,  2013, \apj, 762, 135

\bibitem[\protect\citeauthoryear{{Sippel} \& {Hurley}}{{Sippel} \&
  {Hurley}}{2013}]{SH2013}
{Sippel} A.~C.,  {Hurley} J.~R.,  2013, \mnras, 430, L30

\bibitem[\protect\citeauthoryear{{Spitzer} L}{{Spitzer}}{1987}]{S1987}
{Spitzer} L J.,  1987, Dynamical evolution of globular clusters.
Princeton University Press

\bibitem[\protect\citeauthoryear{{Spitzer} Jr.}{{Spitzer}}{1969}]{S1969}
{Spitzer} Jr. L.,  1969, \apjl, 158, L139

\bibitem[\protect\citeauthoryear{{Spitzer} Jr. \& {Hart}}{{Spitzer} \&
  {Hart}}{1971}]{SH1971}
{Spitzer} Jr. L.,  {Hart} M.~H.,  1971, \apj, 164, 399

\bibitem[\protect\citeauthoryear{{Statler}, {Ostriker} \& {Cohn}}{{Statler}
  et~al.}{1987}]{SOC1987}
{Statler} T.~S.,  {Ostriker} J.~P.,    {Cohn} H.~N.,  1987, \apj, 316, 626

\bibitem[\protect\citeauthoryear{{Trenti}, {Vesperini} \& {Pasquato}}{{Trenti}
  et~al.}{2010}]{TVP2010}
{Trenti} M.,  {Vesperini} E.,    {Pasquato} M.,  2010, \apj, 708, 1598

\bibitem[\protect\citeauthoryear{{Vesperini}, {McMillan} \& {Portegies
  Zwart}}{{Vesperini} et~al.}{2009}]{VMPZ2009}
{Vesperini} E.,  {McMillan} S.~L.~W.,    {Portegies Zwart} S.,  2009, \apj,
  698, 615

\bibitem[\protect\citeauthoryear{{von Hoerner}}{{von Hoerner}}{1957}]{vH1957}
{von Hoerner} S.,  1957, \apj, 125, 451

\bibitem[\protect\citeauthoryear{{Weinberg}}{{Weinberg}}{1994a}]{W1994a}
{Weinberg} M.~D.,  1994a, \aj, 108, 1398

\bibitem[\protect\citeauthoryear{{Weinberg}}{{Weinberg}}{1994b}]{W1994b}
{Weinberg} M.~D.,  1994b, \aj, 108, 1403

\bibitem[\protect\citeauthoryear{{Whitehead}, {McMillan}, {Vesperini} \&
  {Portegies Zwart}}{{Whitehead} et~al.}{2013}]{WMVPZ2013}
{Whitehead} A.~J.,  {McMillan} S.~L.~W.,  {Vesperini} E.,    {Portegies Zwart}
  S.,  2013, ArXiv e-prints

\bibitem[\protect\citeauthoryear{{Zonoozi}, {K{\"u}pper}, {Baumgardt}, {Haghi},
  {Kroupa} \& {Hilker}}{{Zonoozi} et~al.}{2011}]{ZKBH2011}
{Zonoozi} A.~H.,  {K{\"u}pper} A.~H.~W.,  {Baumgardt} H.,  {Haghi} H.,
  {Kroupa} P.,    {Hilker} M.,  2011, \mnras, 411, 1989

\end{thebibliography}

\appendix
\section{Additional Comparison against simulations}
\label{ap:extras}

\begin{figure*}
\centering
\includegraphics[width=160mm]{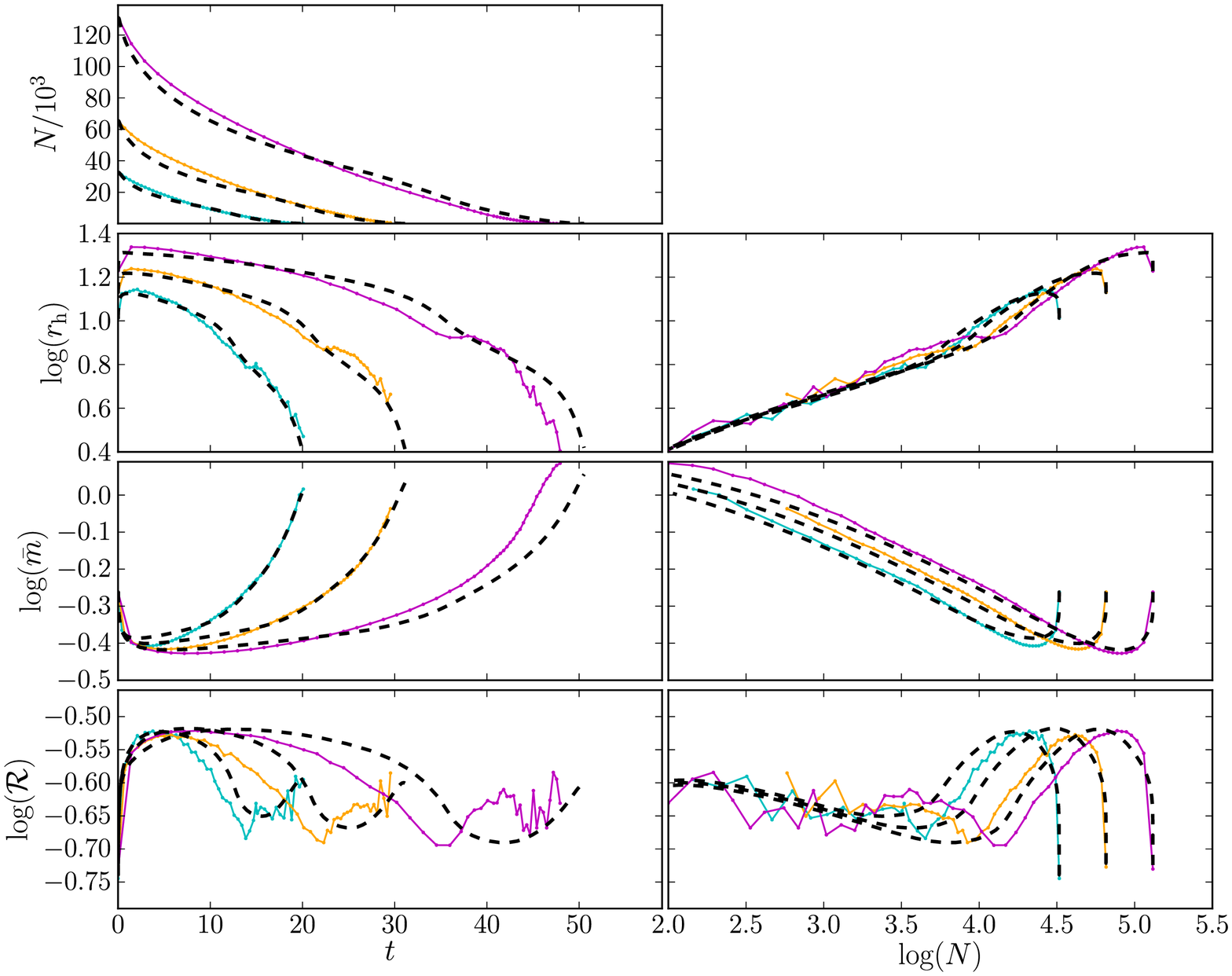}
\caption{Comparison of the \textsc{emacss} model described in sections~\ref{s:sources}, \ref{s:sinks1} and~\ref{s:sinks2} to $N$-body data for a series of RV filling $N$-body simulations of clusters in tidal fields. The clusters are located within an isothermal galaxy halo with $\vg = 220\; {\kms}^{-1}$, at $\rg = 15$ kpc. The left column plots $N$, $\rh$, $\mm$, and $\rhrj$ as a function of time, while the right column plots the same properties as a function of $\log(N)$ .The initial conditions of the $N$-body simulations are described in section \ref{s:nb}, with $N_0$=128k (cyan track; longest surviving), $N_0$=64k (magenta), $N_0$=32k (orange) stars. The clusters are allowed to evolve until their eventual dissolution. For these SCs, $\rhrjr \simeq 0.19$. The observed discrepancies are most likely effects of the (simplified) definitions of $f_{\rm ind}$ and $\gdis$, which follow from the fact that simulations show the highest RV filling and consequently the fastest induced escape.}
\label{f:15kpc}
\end{figure*}

\begin{figure*}
\centering
\includegraphics[width=160mm]{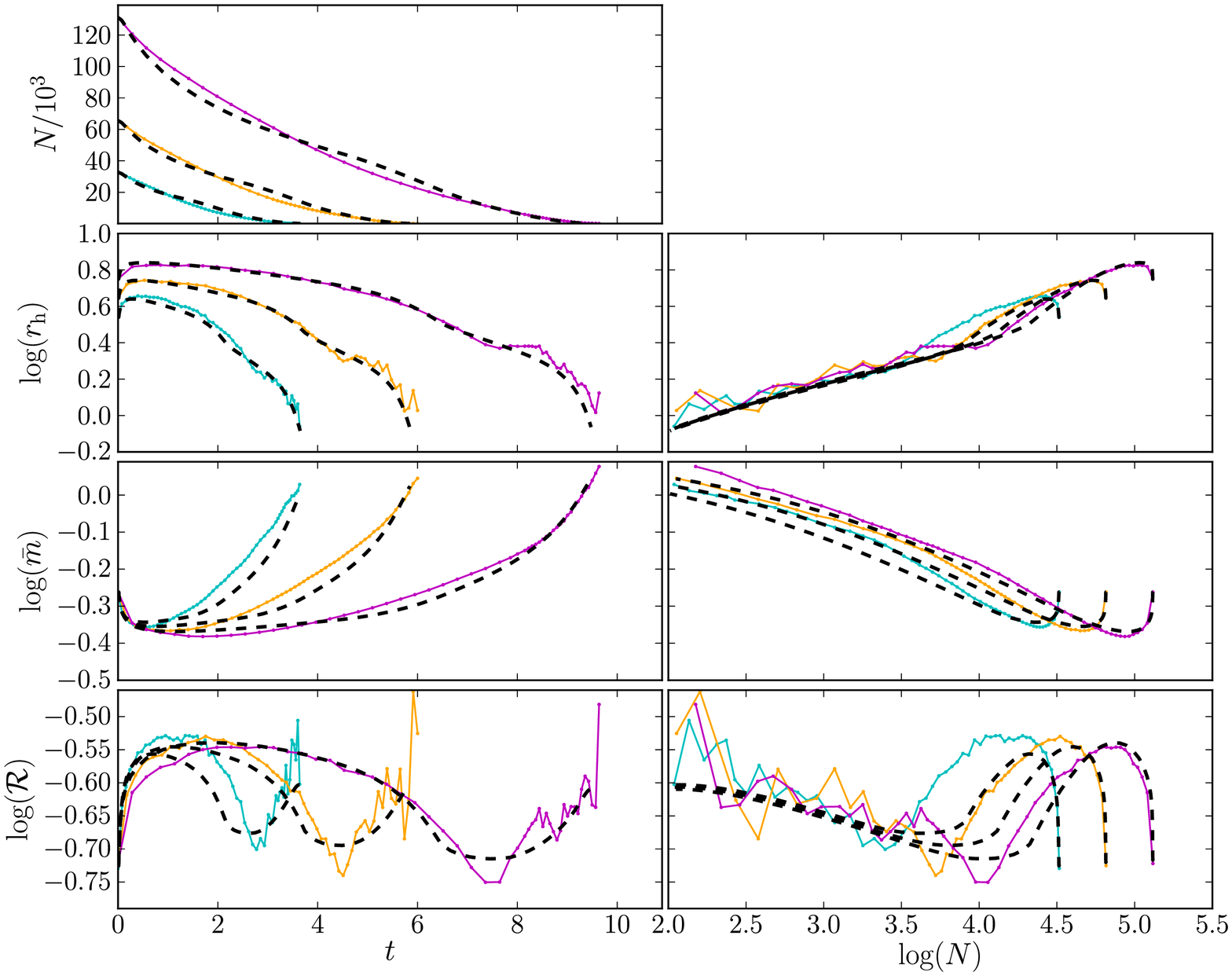}
\caption{Same as Fig.~\ref{f:15kpc} but for clusters located at $\rg = 2.8$ kpc.}
\label{f:2.8kpc}
\end{figure*}

\begin{figure*}
\centering
\includegraphics[width=160mm]{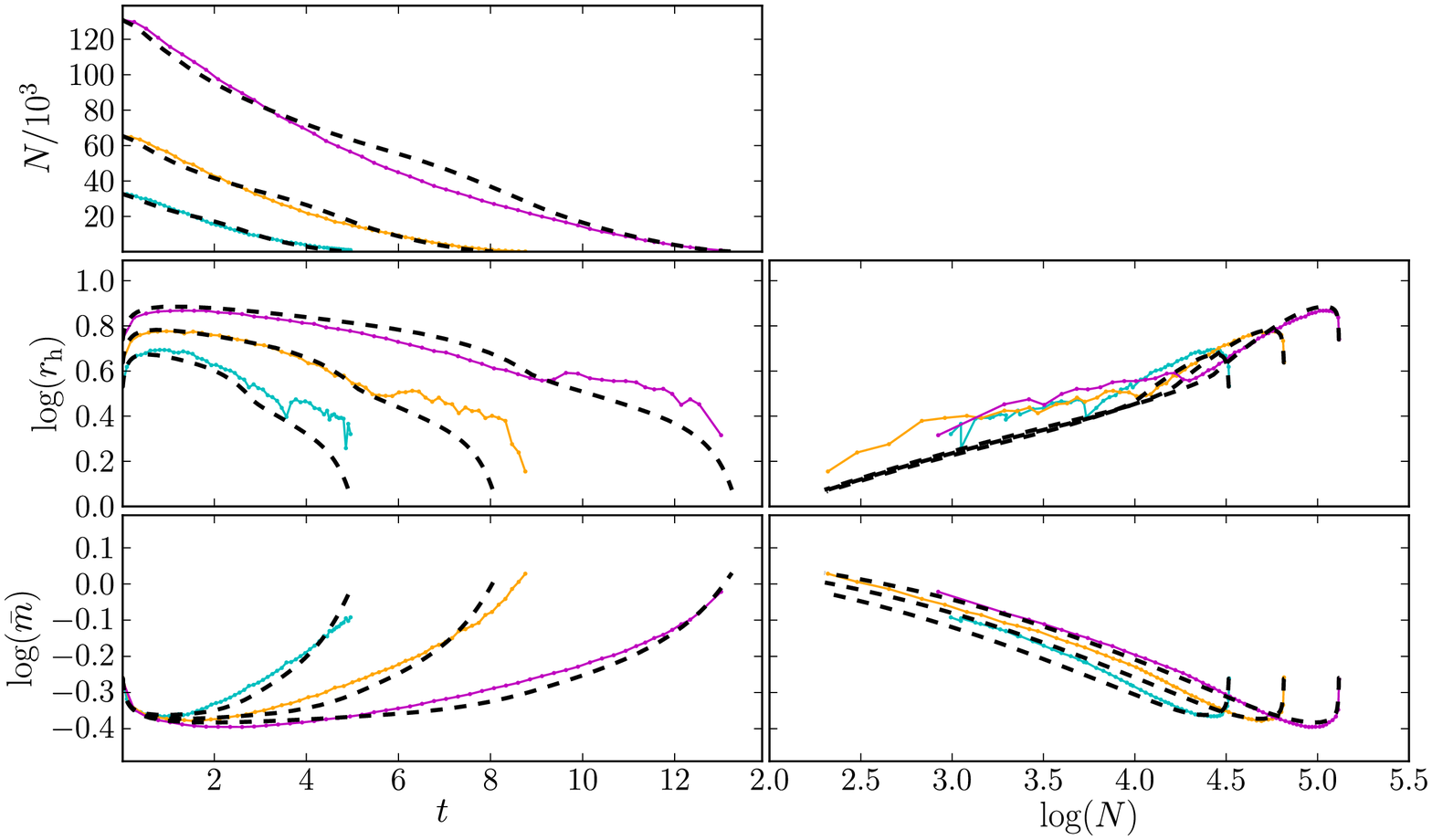}
\caption{Same as Figs.~\ref{f:15kpc} and~\ref{f:2.8kpc} but clusters on $e=0.5$ eccentric orbits with apocentre $\ra = 8.5$ kpc. The three clusters shown have initial $\rh$ chosen to be RV filling, such that $\rhrjr \simeq 0.19$. In these simulations, we have used the approximation of \citet{BM2003}, and have therefore approximated the eccentric orbit as a circular orbit at radius $\rg = \ra(1-e)$ (see section~\ref{s:ecc}). We find that \textsc{emacss} predicts mass-loss and total lifetime extremely well, although during the balanced phase the half-mass radius is under-predicted by a factor $\lesssim 2$. {While this factor is comparable with the the case of a circular orbit, we find increasing discrepancies between $N$-body simulation of \textsc{emacss} prediction for higher eccentricities (e.g. $\gtrsim 2$ in $\rh$, or $\simeq1000$ in $N$ for $e=0.8$). However, such uncertainty is comparable with the assumptions made for statistical studies of observational SCs \citep[e.g. see][]{SKYK2013,AG2013}}. We suggest that the under-prediction of radius occurs since the clusters spend a substantially greater fraction of their lifetime near apocentre, where the cluster can expand further than predicted by our circular orbit approximation. As the cluster passes through apocentre, stars are stripped away (hence leading to our correct mass-loss description), while the rapidly varying tidal field strength causes adiabatic shocking in the cluster \citep[e.g. see][]{W1994a,W1994b}. The consequence of this injection of energy is expansion, according to equation~(\ref{eq:mu}), with comparatively few escapers due to the weaker tidal field.}
\label{f:ecc}
\end{figure*}

In this appendix we show the comparison of $N$-body data from simulated clusters presented by \citet{BM2003} against the equivalent (predicted) evolution of \textsc{emacss}. The clusters simulated are located at $\rg = 15$ kpc (Fig.~\ref{f:15kpc}) or $\rg = 2.8$ kpc (Fig.~\ref{f:2.8kpc}), and are all initially RV filling in the same way as those in Fig.~\ref{f:tfitRF}. Finally, we show simulated clusters and equivalent evolutionary tracks for clusters on $e = 0.5$ eccentric orbits with apocentric distance $\ra = 8.5$ kpc, in figure Fig.~\ref{f:ecc}.

\end{document}